\documentclass[a4paper]{article}
\pdfoutput=1

\usepackage[english]{babel}
\usepackage[utf8x]{inputenc}
\usepackage[T1]{fontenc}

\usepackage[a4paper,top=3cm,bottom=2cm,left=3cm,right=3cm,marginparwidth=1.75cm]{geometry}

\usepackage{amsmath}
\usepackage{mathtools}

\usepackage{graphicx}
\usepackage[colorinlistoftodos]{todonotes}
\usepackage{xcolor}
\usepackage{bm}
\usepackage{lscape}
\usepackage[flushleft]{threeparttable}
\usepackage{lscape}
\usepackage{amsmath}
\usepackage{multirow}
\usepackage[capposition=top]{floatrow}
\usepackage{graphicx}
\usepackage{subcaption}

\usepackage[colorinlistoftodos]{todonotes}
\usepackage[colorlinks=true, allcolors=blue]{hyperref}

\newcommand{\indep}{\rotatebox[origin=c]{90}{$\models$}}

\usepackage{authblk}

\usepackage{setspace}

\author[1]{Kelly Van Lancker}
\author[2]{An Vandebosch}
\author[1, 3]{Stijn Vansteelandt}

\affil[1]{Department of Applied Mathematics, Computer Science and Statistics, Ghent University, Ghent, Belgium}
\affil[2]{Janssen R\&D, a division of Janssen Pharmaceutica NV, Beerse, Belgium}
\affil[3]{Department of Medical Statistics, London School of Hygiene and Tropical Medicine, London, United Kingdom}

\title{\Large Improving interim decisions in randomized trials by exploiting information on short-term outcomes and prognostic baseline covariates}
\begin{document}
\maketitle
\nocite{Tsiatis2006, Halperin1982, Lachin2005, Stallard2010, HampsonJennison2013, Niewczas2016, Austin2010, Assmann2000,
Pocock2002, Qian2016, Bauer2006, Bretz2009, Bauer1989,
Bauer1994, Lehmacher1999, Scharfstein1997, Scharfstein1997, LanWittes1988, Yang2001, Moore2009, proschan2006, Tsiatis2008, Prentice1989}
\begin{abstract}
Conditional power calculations are frequently used
to guide the decision whether or not to stop a trial for futility or to modify planned
sample size. These ignore the information in short-term endpoints and baseline covariates, and thereby do not make fully efficient use of the information in the data. We therefore propose an interim decision procedure based on the conditional power approach which exploits the information contained in baseline covariates and short-term outcomes. We will realise this by considering the estimation of the treatment effect at the interim analysis as a missing data problem. 
This problem is addressed by employing specific
prediction models for the long-term endpoint which enable the incorporation of baseline covariates and multiple short-term endpoints. 
We show that the proposed procedure leads to an efficiency gain and a reduced sample size, without compromising the Type I error rate of the procedure, even when the adopted prediction models are misspecified. In particular, implementing our proposal in the conditional power approach allows earlier decisions relative to standard approaches, whilst controlling the probability of an incorrect decision. This time gain results in a lower expected number of recruited patients in case of stopping for futility, such that fewer patients receive the futile regimen. We explain how these methods can be used in adaptive designs with unblinded sample size reassessment based on the inverse normal $p$-value combination method to control type I error. We support the proposal by Monte Carlo simulations based on data from a real clinical trial.
\end{abstract}
\section{Introduction}
Statistical rules to guide the decision of whether or not to stop a clinical trial early for futility and to adapt the sample size are often based on the conditional power (e.g.\, Halperin et al.\,,$1982$; Lachin, $2005$). 
This monitoring approach quantifies the probability of rejecting the null hypothesis at the end of the study based on a chosen statistical test, given the primary endpoint data observed thus far and an assumption about the future primary endpoint data (e.g.\, the effect size used when powering the study). 
While this methodology is easy to implement and typically well understood by the clinical team, it ignores information in short-term endpoints and baseline variables that can improve the precision of treatment effect estimates. Such increased precision implies a higher probability of stopping early for futility in the absence of a treatment effect, a reduction in average sample size after sample size reassessment and a gain in time. 
 
In view of the above, recent research has focused on incorporating predictive baseline covariates (Qian et al.\,, $2018$) and short-term measurements in interim analyses (e.g.\, Stallard, $2010$; Hampson and Jennison, $2013$; Niewczas et al.\,, $2016$). This brings several challenges. First, the additional information accrued between the interim analysis and the final analysis may involve observations of subjects who already contributed information to the interim analysis. 
This makes it more challenging to maintain whether the independent increments property of the Brownian motion structure, used to justify the conditional power approach (Lachin, $2005$). 
Second, the incorporation of baseline covariates and short-term endpoints requires the postulation of statistical models. This raises concerns that their misspecification may result in bias (Austin et al.\,, $2010$; Assmann et al.\,, $2000$; Pocock et al.\,, $2002$). 

In this paper, we will overcome the abovementioned two concerns. For this, the precision of the treatment effect estimator at the interim analysis will be optimized by incorporating short-term endpoints as well as baseline covariates. 
We will realise this by considering the estimation of the treatment effect at the time of the interim analysis as a missing data problem. This problem can be addressed by making use of specific prediction models for the long-term endpoint which, besides multiple short-term endpoints, can also incorporate baseline covariate information. 
This methodology surprisingly has the appealing feature that it delivers a consistent treatment effect estimator.
Of all estimators that have this property, the considered one is most efficient (provided that all models are correct).
The proposed estimator is closely related to one given in the appendix of Qian et al. ($2016$). 
In contrast to Qian et al. ($2016$), we justify the proposed procedure to imbed the interim test statistic based on this proposed interim estimator in the conditional power approach by relating to the general work of Scharfstein et al.\,($1997$) on an information based-design and monitoring procedure.
This extension of the conditional power approach allows earlier stopping for true futility whilst controlling the probability of incorrectly stopping. 
This generally leads to a reduction in the number of recruited patients in the case of stopping for futility, such that fewer patients receive the futile regimen.

Moreover, to further allow for modifying the design of the remaining study if the trial is not stopped early for futility, we next extend the method to adaptive designs with unblinded sample size reassessment based on conditional power arguments (Bauer and K\"onig, $2006$). Since sample size adaptations can inflate the type I error (Bretz et al.\,, $2009$), we use the adaptive combination test proposed by Bauer ($1989$) and Bauer and K\"ohne ($1994$) based on the inverse normal method as the combination function (Lehmacher and Wassmer, 1999). We also justify the independence of the stage-wise test statistics used in the combination test by relating to the general work in Scharfstein et al.\,($1997$). 
We support the proposal with simulation studies based on data from a real clinical trial.
\subsection{Motivating Example}
The motivating phase $3$ clinical trial was designed to evaluate the efficacy of a new experimental treatment for multidrug-resistant tuberculosis on top of the standard of care regimen (referred to as background regimen BR) as compared to placebo plus BR with regard to the proportion of subjects with a favorable treatment outcome (ClinicalTrials.gov, NCT00449644 \cite{Diacon}) defined as confirmed culture conversion $60$ weeks after randomization. 
Besides the clinical endpoint of interest, (predictive) baseline covariates as well as confirmed culture conversion (cure) $16$ weeks after randomization were planned to be measured. The available prior phase $2$b data suggested that the early measurements are reasonably predictive of the the long-term measurements, in the sense that for the majority of subjects achieving the primary endpoint (culture conversion at week $60$), culture conversion is expected to have occurred by week $16$. 

An interim analysis was planned to be performed prior to enrollment completion to evaluate whether the data collected up to that point contain any evidence for superiority and if not, the trial would be stopped for futility. Since many enrolled patients may have no information on the primary endpoint available at the time of the interim analysis, restricting the interim analysis to those patients with long-term information available may result in lack of information to support futility decisions. To have a reasonable chance of detecting true futility early, we will include information on baseline covariates and short-term endpoints into interim analyses of the long-term endpoint, and thereby make fully efficient use of the information in the data. Guided by this motivating example, the paper will focus on binary endpoints. However, the proposed approach is applicable to other endpoints (e.g.\, continuous and survival endpoints).
\section{Setting and Estimation}
\subsection{Trial Design}\label{setting1}
Consider a study design which intends to collect i.i.d.\,data \{($Y_i$, $X_i$, $\bm{Z}_i$, $A_i$), $i=1, \dots, n$\}, with $Y_i$ the binary primary endpoint, $X_i$ a correlated short-term endpoint and $\bm{Z}_i$ a vector of prognostic baseline covariates (e.g. age, gender, \dots) for patient $i$ who is randomly assigned to either experimental treatment ($A_i = 1$) or control ($A_i = 0$).
For pedagogic  purposes, we consider a single short-term endpoint $X$, but all results extend directly to the general case with multiple short-term endpoints.   
Let $n_j$ ($j\in\{0,1\}$) correspond to the pre-planned number of patients in each treatment arm and $n=n_1+n_0$ the planned sample size.
Define $P_j$ ($j\in\{0,1\}$) as the probability of a successful outcome at the end of a trial in the experimental ($j=1$) and control ($j=0$) arm. The primary hypothesis of interest, $H_0: P_1 = P_0$, will be tested against the one-sided alternative, $H_A: P_1 > P_0$ at level $\alpha$ with power $1-\beta$. 
To evaluate this hypothesis, the test statistic for the difference in marginal sample proportions of the two arms is compared to $z_{1-\alpha}$: 
\begin{align*}
Z=\frac{\hat{P}_1-\hat{P}_0}{\sqrt{\frac{\hat{P}_1(1-\hat{P}_1)}{n_1}+\frac{\hat{P}_0(1-\hat{P}_0)}{n_0}}},
\end{align*}
where $\hat{P}_1=\frac{\sum_{i=1}^{n}A_iY_i}{\sum_{i=1}^{n}A_i}$ and $\hat{P}_0=\frac{\sum_{i=1}^{n}(1-A_i)Y_i}{\sum_{i=1}^{n}(1-A_i)}$ denote the estimated probability of a successful outcome at the end of the trial in respectively the experimental and control arm respectively.
\subsection{Effect Estimates at the Interim Analysis}\label{section_estimator}
Consider an interim analysis of the primary endpoint $Y$ and assume that patients are continuously recruited during the course of the trial. Since not all patients have full data observed at the interim analysis, we use $C^Y_i$ to denote whether $Y_i$ is already observed ($C^Y_i=1$) or not ($C^Y_i=0$), $C^X_i$ to denote whether $X_i$ is already observed ($C^X_i=1$) or not ($C^X_i=0$) and $C^Z_i$ to denote the missingness status of $\bm{Z}_i$: $C^Z_i=1$ if and only if $\bm{Z}_i$ is observed (i.e.\,if and only if patient $i$ has been enrolled in the study at the time of the interim analysis). The estimation of the effect at an interim analysis can then be seen as a missing data problem, where we can distinguish $4$ cohorts of patients: 
a first cohort of patients for whom all data are available (($C^Z_i$, $C^X_i$, $C^Y_i$)=($1$, $1$, $1$)), a second cohort of patients who passed the timepoint at which $X$ is evaluated but not yet $Y$ and thus for whom only $\bm{Z}$ and $X$ are observed (($C^Z_i$, $C^X_i$, $C^Y_i$)=($1$, $1$, $0$)), a third cohort of patients for whom only $\bm{Z}$ is observed (($C^Z_i$, $C^X_i$, $C^Y_i$)=($1$, $0$, $0$)) and a fourth cohort of patients who have not yet been recruited (($C^Z_i$, $C^X_i$, $C^Y_i$)=($0$, $0$, $0$)). 
Although the end-of-study outcome under the experimental treatment is only seen for the treated patients in cohort $1$, under the assumption that recruitment occurs randomly 
(i.e.\,independent censoring holds, in the sense that ($C^Z_i$, $C^X_i$, $C^Y_i$) is independent of $\bm{Z}_i$, $A_i$ and the potential outcomes of $Y_i$ and $X_i$), the missing outcomes for the other recruited patients on treatment can be unbiasedly predicted in large samples (Tsiatis, $2006$). Similar to the closely related estimator described in the appendix of Qian et al.\, ($2016$), this can de done as follows:

\begin{enumerate}
		\item regress $Y$ on $X$ and $\bm{Z}$ among the patients in cohort $1$ of the treatment arm ($C^Y=1$ and $A=1$) using a canonical generalized linear working model for the conditional mean of $Y$: $E(Y|A=1, X, \bm{Z})=h_1(X, \bm{Z}, \bm{\eta}_0)$, where $h_1(X, \bm{Z}, \bm{\eta})$ is a known function, evaluated at a parameter $\bm{\eta}$ with unknown population value $\bm{\eta}_0$; for example $h_1(X, \bm{Z}, \bm{\eta})=\text{logit}^{-1}( \eta_1+\eta_2X+\bm{\eta}_3\bm{Z})$,
	\item use this regression model to predict the outcome under the experimental treatment for the treated patients in cohort $2$ ($C^X=1$, $A=1$ and $C^Y=0$) based on their observed baseline covariates $\bm{Z}$ and short-term endpoint $X$ as $\hat{Y}_{1i}=h_1(X_i, \bm{Z}_i, \hat{\bm{\eta}})$,
	\item then, regress the combination of these fitted values for the treated patients in cohort $2$ and the observed $Y$ values for the treated patients in cohort $1$ ($Y^*:=C^YY+(1-C^Y)C^X\hat{Y}_{1}$) on the baseline covariates $\bm{Z}$ using a canonical generalized linear working model for the conditional mean of $Y^*$: $E(Y^*|A=1, \bm{Z})=f_1(\bm{Z}, \bm{\delta}_0)$, where $f_1(\bm{Z}, \bm{\delta})$ is a known function, evaluated at a parameter $\bm{\delta}$ with unknown population value $\bm{\delta}_0$; for example $f_1(\bm{Z}, \bm{\delta})=\text{logit}^{-1}( \delta_1+\bm{\delta}_2\bm{Z})$,  
	\item and use this regression model to predict $Y$ for the treated patients in cohort $3$ based on their observed baseline covariates $\bm{Z}$ as $\hat{Y}_{1i}'=f_1(\bm{Z}_i, \hat{\bm{\delta}})$.
\end{enumerate}
An interim estimator $\hat{\mu}_{1}$ of $P_1$ is obtained by taking the average of the observed $Y$-values for the treated patients in cohort $1$, the predicted values $\hat{Y}_{1i}$ based on $X_i$ and $\bm{Z}_i$ for the treated patients in cohort $2$ and the predicted values $\hat{Y}_{1i}'$ based on $\bm{Z}_i$ for the treated patients in cohort $3$. 
A theoretical derivation of this estimator and its properties are given in Appendix A.
The interim estimator $\hat{\mu}_{0}$ for the outcome under control can be analogously estimated by applying the same steps to the untreated patients $A=0$.
If the study design does not intend to collect an earlier measurement $X$, the first two steps can be omitted since there are then no patients in cohort $2$. $Y^*$ in step $3$ then reduces to $C^YY$; this then involves fitting a model for $Y$ given $\bm{Z}$ among the (un)treated patients in cohort $1$.

As a result of the random recruitment and simple randomisation, the estimators $\hat{\mu}_{1}$ and $\hat{\mu}_{0}$ have the appealing feature that misspecification of the outcome models in step $1$ and $2$ does not introduce bias in large samples (see Appendix A.2). 
Moreover, when the outcome models are correctly specified, these estimators are asymptotically efficient in the subclass of estimators that are unbiased as soon as ($A$, $C^Z$), $C^X$ and $C^Y$ are independent of respectively $\bm{Z}$, ($\bm{Z}$, $X$) and ($\bm{Z}$, $X$, $Y$) (e.g.\,, Yang and Tsiatis, 2001; Tsiatis, 2006; Moore and van der Laan, 2009; Stallard, 2010; Qian, 2018). 
 
In order to calculate the variance of the treatment effect $\hat{\mu}_{1}-\hat{\mu}_{0}$, one must take into account that the predictions are estimated based on particular outcome regression models.
It is therefore not sufficient to compute the variance of the predictions and observed values of the primary endpoint.
In Appendix A.$3$, we show that this can be easily accommodated under randomization and random recruitment by calculating the asymptotic variance of $\hat{\mu}_{1}-\hat{\mu}_{0}$, denoted $s^2$, as $1/n'$ times the sample variance of the values
\begin{align}\label{variance_end}
\begin{split}
&(A_i/\hat{\pi})[\{C^Y_iC^X_i/(\hat{\pi}^Y\hat{\pi}^X)\}(Y_i-\hat{Y}_{1i})+(C^X_i/\hat{\pi}^X)(\hat{Y}_{1i}-\hat{Y}_{1i}')+\hat{Y}_{1i}'-\hat{\mu}_{1}] \\
-&((1-A_i)/(1-\hat{\pi}))[\{C^Y_iC^X_i/(\hat{\pi}^Y\hat{\pi}^X)\}(Y_i-\hat{Y}_{0i})+(C^X_i/\hat{\pi}^X)(\hat{Y}_{0i}-\hat{Y}_{0i}')+\hat{Y}_{0i}'-\hat{\mu}_{0}],
\end{split} 
\end{align}
with $\hat{\pi}=\hat{P}(A=1)$ the observed randomization probability, $\hat{\pi}^X=\hat{P}(C^X=1)$ and $\hat{\pi}^Y=\hat{P}(C^Y=1|C^X=1)$ and $n'$ denoting the total number of recruited patients at the time of the interim analysis.
We refer the interested reader to Appendix A.3 for more details. 
\subsubsection{Precision Gain based on Baseline Covariates and Short-Term Endpoints}
As mentioned before, in large samples, our proposed estimator is never outperformed by the standard analyses that use only information on the short-term and/or primary endpoint. The proposed estimator also reduces to these standard analyses if no baseline covariates are available.
The magnitude of the precision gain itself depends on different characteristics of the interim analysis and available data. 
First, the predictivity of the short-term endpoints and baseline covariates plays a very important role: the more predictive they are, the more information available, and thus, the larger the gain in efficiency relative to respectively the standard analyses that use only information on the primary endpoint and standard analyses that use only information on the primary and short-term endpoint. 
Nonetheless, adjusting for a prognostic baseline covariate alone always leads to larger precision gain than adjusting for an equally prognostic short-term outcome alone because of the larger number of partially observed patients (cohort $1$ and $2$) compared to the number of patients in cohort $2$ only (Qian et al\,, $2016$). 
Second, the recruitment rate and the time at which the interim analysis is conducted, are also of great value since they influence the number of patients in each cohort. 
For example, the relative benefit that results from incorporating baseline covariates and short-term endpoints is generally smaller at later interim analyses since more primary endpoint data $Y$ are then available.
Lastly, the time at which the short-term endpoint is measured, is crucial. Assuming two equally predictive short-term endpoints, adjusting for the earliest measured endpoint will be more advantageous since it will be measured for more patients relative to adjusting for the one measured later in time.
\section{Interim Analysis}
\subsection{Futility stopping Based on Conditional Power}
To calculate conditional power, we need to define how far through the trial we are at the time of the interim analysis. This can be expressed in terms of the information fraction $t$, which is defined as the fraction of information available at the time of the interim analysis versus the expected information at the end of the study; i.e.\,the fraction of the expected variance of the treatment effect estimator at the end of the study versus the variance of the interim estimator. Define $Z_t$ and $Z_1$ as the $Z$-statistic of the treatment effect at information fraction $t$ and at the end of the trial, respectively, and let $A\indep B$ for random variables $A$ and $B$ denote that $A$ is independent of $B$.
 For any test satisfying the independent increments property, i.e.\,$\sqrt{t}Z_t \indep (Z_1-\sqrt{t}Z_t)$, the conditional power can be computed using the Brownian motion structure in combination with the $B$-value (Lan and Wittes, $1988$), which is defined as $B_t=Z_t\sqrt{t}$, $0<t\leq1$:
\begin{align*}
CP_t(\theta) = 1-\Phi\left(\frac{z_{1-\alpha}-Z_t\sqrt{t}-\theta(1-t)}{\sqrt{1-t}}\right),
\end{align*}
with $\theta$ the assumed drift parameter. This parameter reflects the expected $Z$-score at the final analysis based on the assumption made about the data to be observed in the remainder of the study, e.g.\,$\theta=z_{1-\alpha}+z_{1-\beta}$ when the effect size used for powering the study is used as assumption for the remaining (unobserved) primary endpoint data. To decide whether or not to stop for futility, the conditional power is compared to some threshold. If the conditional power is below this threshold, the trial is stopped for futility; otherwise the trial is continued. 

The traditional conditional power approach uses a test statistic based on the treatment-specific sample averages at the time of the interim analysis
$
\sum_{i=1}^{n}C^Y_iA_iY_i/\sum_{i=1}^{n}C^Y_iA_i$
and 
$
\sum_{i=1}^{n}C^Y_i(1-A_i)Y_i/\sum_{i=1}^{n}C^Y_i(1-A_i)$, which only use the complete cases (i.e., $C^Y_i=1$). It seems appealing to employ instead the interim test statistic based on the more efficient estimator introduced in the previous section
\begin{align*}
Z_t=\frac{\hat{\mu}_{1}-\hat{\mu}_{0}}{\sqrt{s^2}},
\end{align*}
which also incorporates information on the short-term outcome and the baseline covariates. Note that the test statistic for the primary analysis (at the end of the study), $Z_1$, coincides with the standard $Z$ test statistic which only incorporates information on the primary endpoint. Since the conditional power is computed using standard Brownian motion arguments, the independent increments property needs to be satisfied. 
A potential concern is that the additional information accrued between the interim analysis and the final analysis involves additional observations of subjects who already contributed information to the interim analysis.
Building on Scharfstein et al.\, ($1997$), we show in Appendix B that the independent increments assumption is nonetheless (asymptotically) maintained when the prediction models are correctly specified since our test is then semiparametric efficient. Interestingly, we show in Appendix B that this is even true when the prediction models are misspecified, assuming that the treatment is randomly assigned and the recruitment is random.

The information fraction $t$ is now calculated as the fraction of the variance of the estimator at the end of the study and the variance $s^2$ of the interim estimator $\hat{\mu}_{1}-\hat{\mu}_{0}$. This is not readily available since at the time of the interim analysis, the variance of the final estimator is unknown as $Y$ is only available for the patients in cohort $1$ ($C^Y_i=1$). 
Nevertheless, under the assumption of random recruitment, $Y\indep C^Y$, the variance of the final estimator can be estimated using only the complete cases since they then form a representative sample. This allows estimating the variance as $1/n$ times the sample variance of $(A_i/\hat{\pi})(Y_i-\hat{\mu}_1)-\{(1-A_i)/(1-\hat{\pi})\}(Y_i-\hat{\mu}_0)$ in cohort $1$.
\subsection{Sample Size Reassessment Based on Conditional Power}
When the trial is not stopped early for futility at the interim analysis, one may still consider modifying the sample size based on the unblinded treatment effect. 
After sample size adaptations, the usual test statistic at the end of the study, which simply pools the outcome data obtained before and after the sample size adjustment, cannot be applied since its naive use might inflate the type I error rate (Bretz et al.\,, $2009$). 
The adaptive $p$-value combination test proposed by Bauer ($1989$) and Bauer and K\"ohne ($1994$) allows to combine the independent test statistics $Z^{(1)}$ and $Z^{(2)}$ based on the data observed before (stage $1$) and after (stage $2$) the interim analysis, respectively, while guaranteeing Type I error rate control. Moreover, this two stage procedure corresponding with the combination test provides a simple
way for sample size reassessment.

The combination test for the final analysis can be obtained with the weighted
inverse normal combination function (Lehmacher and Wassmer, 1999), which can be written as
\begin{align*}
C(Z^{(1)}, Z^{(2)}) = 1 - \Phi(\sqrt{w}Z^{(1)}+\sqrt{1-w}Z^{(2)}),
\end{align*}
where $\Phi$ denotes the standard normal cumulative distribution function, and $w$ a pre-specified weight which can be chosen arbitrarily between $0$ and $1$. It follows from Lehmacher and Wassmer ($1999$) that, no matter the choice of weight $w$, rejecting the null hypothesis when $C(Z^{(1)}, Z^{(2)})<\alpha$, leads to a test at level $\alpha$ when $Z^{(1)}$ and $Z^{(2)}$ are independent.
A natural choice for the weight $w$ is the originally planned information fraction $t$ at which the interim analysis is performed, since the inverse normal combination test using normally distributed test statistics then coincides with a group sequential design test statistic at the final analysis when sample size remains as pre-planned \cite{wassmer2016}. 
The interim test statistic $Z_t$ is used as the test statistic corresponding with the first stage, since it is based on all data observed before the interim analysis.
Now, define $\tilde{Z}_1$ as the naive standard (unweighted) test statistic for the primary endpoint at the end of the trial after sample size re-estimation and $\tilde{t}$ as the fraction of information available at the interim analysis relative to the information available at the end of the study after sample size re-assessment; i.e.\,the fraction of the variance of the treatment effect estimator at the end of the trial after sample size re-estimation versus the variance of the interim estimator.
We propose $Z^{(2)}$, the test statistic based on the data to be observed in the remainder of the study, to equal $(\tilde{Z}_1-\sqrt{\tilde{t}}Z_t)/\sqrt{1-\tilde{t}}$, which is asymptotically independent of $Z_t$ (see Appendix B and C.1) and guarantees that $\sqrt{t}Z^{(1)}+\sqrt{1-t}Z^{(2)}$ equals $\tilde{Z}_1$ if there are no adaptations. In particular, the proposed adaptive combination test can be written as
\begin{align*}
C(Z^{(1)}, Z^{(2)}) = 1 - \Phi\left(\sqrt{t}Z_t+\sqrt{1-t}\frac{\tilde{Z}_1-\sqrt{\tilde{t}}Z_t}{\sqrt{1-\tilde{t}}}\right).
\end{align*}

Using this combination test makes it possible to perform a sample size reassessment during the course of the trial. 
The total sample size is adapted so as to make the conditional power equation equal to the pre-specified design power of $1-\beta$, assuming the effect size used for powering the study for the remaining (unobserved) primary endpoint data (Bauer and K\"onig, $2006$). Based on the principle that the a priori fixed weights derived from the planned information fraction at the interim analysis are used for the calculation of the final test-statistics, rearrangement of the formulae of Bauer and K\"onig ($2006$) (see Appendix C.2) leads to
\begin{align}\label{ssr}
\tilde{n}-t\cdot n = \left(\frac{\frac{z_{1-\alpha}-Z_t\sqrt{t}}{\sqrt{1-t}}-z_{\beta}}{\theta/\sqrt{n}}\right)^2,
\end{align} 
where $\tilde{n}$ corresponds to the total sample size after adaptation. 
Note that $\frac{z_{1-\alpha}-Z_t\sqrt{t}}{\sqrt{1-t}}-z_{\beta}<0$ points to the fact that the new experimental treatment is significantly better and no further recruitment is needed. Considering that the total sample size is bounded to equal at least the number of recruited patients at the interim analysis, we then obtain
\begin{align}\label{SS}
\tilde{n} = \max\left(n', \left(\frac{\max\left(0, \frac{z_{1-\alpha}-Z_t\sqrt{t}}{\sqrt{1-t}}-z_{\beta}\right)}{\theta/\sqrt{n}}\right)^2 + t\cdot n\right),
\end{align} 
with $n'$ the total number of patients recruited at the interim analysis. If a sample size decrease is not allowed, the new sample size is obtained by taking the minimum of $n$ and the second part in formula (\ref{SS}).
In practice, sample size reassessment may be done carrying forward the observed treatment effect instead of the effect size used for powering the study, by replacing $\theta$ in the previous formulas by $Z_t/\sqrt{t}$.

\subsection{Blinded Information Fraction}
Usually, the interim analysis is implemented at the time at which the pre-planned information fraction $t$ is reached.
Calculating this information fraction can be either done in a blinded or unblinded way. Unblinding the treatment assignment code, however, requires an independent data monitoring committee to review the accumulating data from the beginning of the trial. We therefore recommend to approach the operational planning alternatively, in a way that preserves blinding. As before, one can calculate the information fraction in a blinded way by estimating the fraction of the variances of the treatment effect estimator at the end of the study versus during monitoring of the trial. The estimates of the two variances should now be based on the blinded data, without making use of the observed data on treatment. 
Under the null hypothesis ($Y_i\indep A_i$) and the assumption that $Y_i\indep C^Y_i$, the variance of the estimator for the treatment effect at the end of the study can be  blindly estimated as $1/(n(1-\hat{\pi})\hat{\pi})$ times the sample mean of $(Y-\hat{\mu})^2$ over all patients in cohort $1$, with $\hat{\mu}=\sum_{i=1}^{n}C^Y_iY_i/\sum_{i=1}^{n}C^Y_i$ (see Appendix D). 
Assuming that $C^X_i \indep Y_i|\bm{Z}_i$ and $C^Y_i \indep Y_i|\bm{Z}_i, X_i, C^X_i=1$, the blinded estimate of $s^2$, the variance of the estimator for the treatment effect during monitoring the trial, is obtained by
following the same steps as in Section \ref{section_estimator} but supposing now that everyone is in the same group. Formulas and their derivation are included in Appendix D.

\section{Simulation Study}

We conducted a simulation study to examine the finite-sample performance of the proposed interim estimator for making early decisions based on the conditional power approach as well as for reassessing the sample size. The impact of the short-term endpoint and baseline covariates on the one hand and model misspecification on the other were investigated. We evaluate the expected total number of subjects, the expected number of subjects needed at the interim analysis, as well as the type $1$ error and power.

\subsection{Data Generating Distributions for Simulation Studies Based on TMC207-C208 Stage 2}
To determine a realistic data generating model, the phase $2$b data of the motivating example were employed. 
This dataset consists of $132$ participants, $66$ in each arm, with observed primary endpoint data. 
We generate a simulated trial of $n$ -depending on the simulation settings- hypothetical participants using the original dataset as follows. We resample with replacement from the original study population of $132$ patients and only extract the baseline covariate information from the subjects in the resampled population. Then, treatment and control are assigned with probability $0.5$ to each hypothetical participant. The short-term endpoint $X$, measured $4$ months after randomization, is predicted based on a Bernoulli distribution with probability determined by a logistic regression model for $X$ fitted on the original phase $2$b dataset. The primary endpoint data, measured $15$ months after randomization, are generated under different scenarios depending on the simulation study. Unless otherwise stated, we consider the same recruitment scenario as in the phase $2$b dataset, where on average $8$ patients per month enter the study. 

Respectively $2,500$ and $10,000$ ($100,000$ for futility scenarios) Monte Carlo simulations were performed for the different settings evaluating the proposal for the conditional power approach and the sample size reassessment. The cut-off values to decide whether or not to stop for futility at the interim analysis are based on the O'Brien-Fleming futility boundaries assuming a total power of $0.90$ and a one-sided $\alpha=0.025$ (see \cite{Zhang2010}).
\subsection{Simulations on Conditional Power}
\subsubsection{Without Short-Term Measurements}\label{sim_without}
First, we conducted a simulation study to show that the conditional power based on the proposed interim estimator outperforms the standard conditional power approach when adjusting for predictive baseline covariates. 
For pedagogic purposes, only one continuous covariate $Z_1$ that might be predictive for cure was selected based on the phase $2$b data. The resulting model -allowing for different degrees of predictivity- was employed to simulate the primary endpoint data for each subject $i$ ($i = 1,\cdots , n$), $Y_i|\bm{Z}	_i\overset{d}{=}Ber\{m_0(\bm{Z}_i)\}$, where  $m_0(\bm{Z})=\text{logit}^{-1}(-0.23+0.75A+c\cdot0.39Z_1+c\cdot0.17Z_1A)$ with $c\in\{0,1,2,3,4,5\}$ and the coefficients $\{-0.23,0.75,0.39, 0.17\}$ correspond to these obtained from a logistic regression of $Y$ on the continuous covariate $Z_1$ fitted on the original data. The proportion of the variance in $Y$ that is predictable from $\bm{Z}=Z_1$ is shown in Table \ref{R2_cp1} (see Appendix E).
\begin{table}[h]
			\centering
	\begin{threeparttable}
		\caption{\label{R2_cp1} Empirical $R$-squared values and probability of succes (under alternative hypothesis) for the data generated under the settings without short-term endpoint. Simulations results are based on $10,000$ Monte Carlo replications.}
		\begin{tabular}{l l  c c c c c c c cc cc}
			\hline	\hline	
			&&\multicolumn{3}{c}{$H_0$}&&\multicolumn{7}{c}{$H_1$}\\
			\cline{3-5} \cline{7-13}\\
			&&\multicolumn{1}{c}{$A=1$}&&\multicolumn{1}{c}{$A=0$}&&\multicolumn{1}{c}{$A=1$}&&\multicolumn{1}{c}{$A=0$}&&&&\\		
			\cline{3-3}  \cline{5-5} \cline{7-7} \cline{9-9} \\
			Model&& $R^2$& & $R^2$ &&  $R^2$& &$R^2$ &&$P_1$ &&$P_0$\\ \hline
			&&& &&  &&& &&&&\\
			$c=0$&& $0.01$& & $0.01$ &&  $0.01$& &$0.01$ &&$0.63$& & $0.44$ \\
			$c=1$&& $0.07$& & $0.07$ &&  $0.12$& &$0.07$ &&$0.62$& & $0.44$ \\
			$c=2$&& $0.20$& & $0.20$ &&  $0.34$& &$0.20$ && $0.60$& & $0.45$\\
			$c=3$&& $0.36$& & $0.36$ &&  $0.54$& &$0.36$  &&$0.57$& & $0.45$\\
			$c=4$&& $0.50 $& & $0.50$ &&  $0.67$& &$0.50$ && $0.55$& & $0.45$ \\
			$c=5$&& $0.61$& & $0.61$ &&  $0.76$& &$0.61$  &&$0.54$& & $0.45$\\
			&&& &&  &&& &&&&\\ \hline
		\end{tabular}
		\begin{tablenotes}
			\small
			\item Note: $R^2$ corresponds with the proportion of the variance in $Y$ that is predictable from $\bm{Z}=Z_1$ (estimated using formulas in Appendix E); $P_1$ and $P_0$, probability of success in respectively the treatment and control arm under superiority.
		\end{tablenotes}
	\end{threeparttable}
\end{table}
The sample size of $227$ patients in each arm corresponds with the required sample size to obtain a power of $0.90$ under a one-sided significance level of $0.025$  for the scenario with $c=2$.
Note that fixing the sample size implies a different power under each setting ($\{0.98,0.97,0.90,0.76,0.61,0.47\}$ for $c=\{0,1,2,3,4,5\}$) since the different scenarios resemble different treatment effects ($P_1-P_0$ in Table \ref{R2_cp1}).

\begin{figure}[h]
		\caption{True Futility: Comparison of the information fraction and the probability to stop for futility between the standard conditional power approach and the conditional power using the proposed interim estimator for different scenarios under true futility. True Superiority: Comparison of the information fraction and the loss of power compared to the design power between the standard conditional power approach and the conditional power using the proposed interim estimator for different scenarios under superiority. The interim analysis is conducted at a fixed point in time.}
	\label{sim_CP_without}
\begin{subfigure}[b]{\textwidth}\caption{\label{sim_CP_without_a} True Futility} 
		\includegraphics[width=\linewidth]{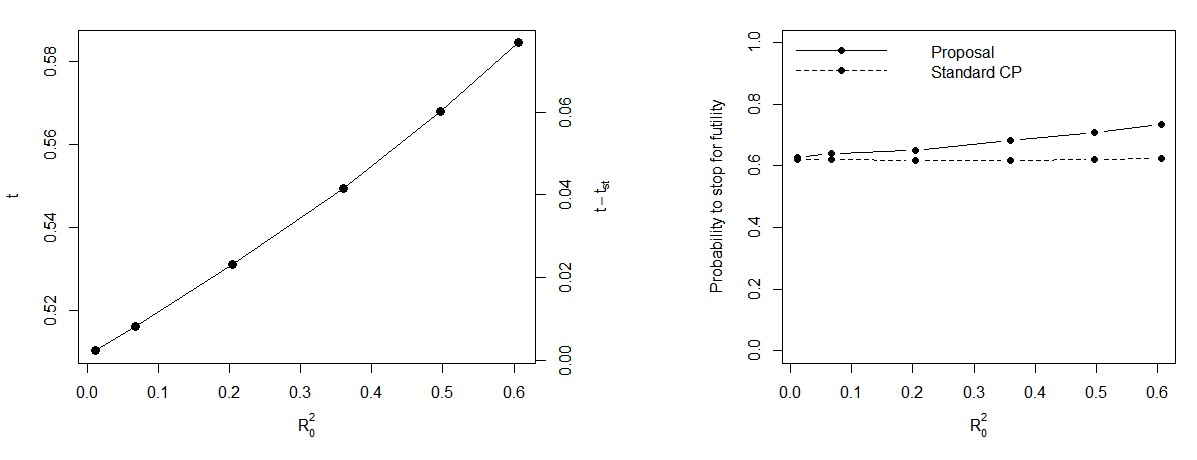}
	\end{subfigure}
\begin{subfigure}[b]{\textwidth}\caption{\label{sim_CP_without_b} Superiority} 
	\includegraphics[width=\linewidth]{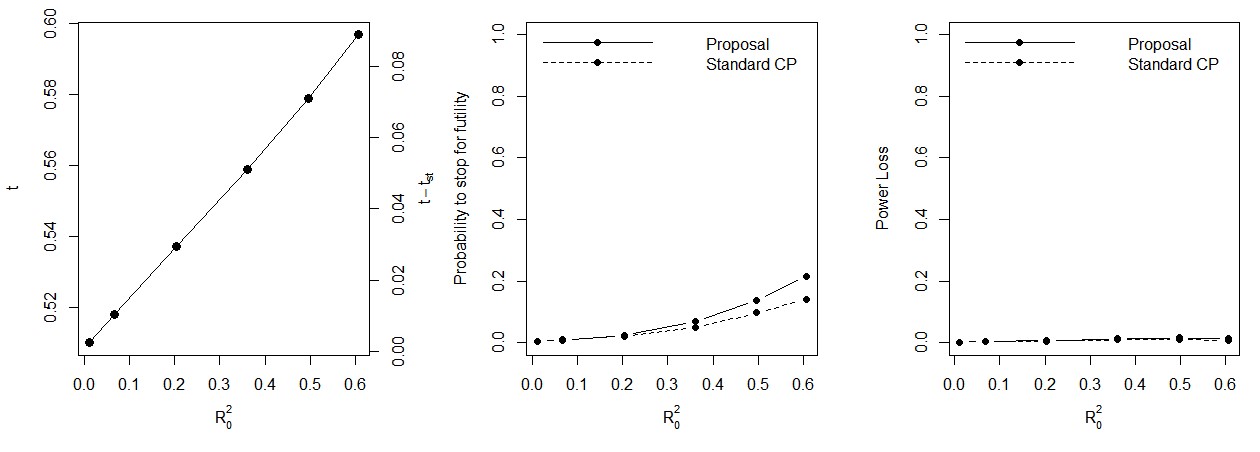}
\end{subfigure}
\floatfoot{
Note: $t$ and $t_{st}$, average information fraction using respectively the proposed and usual test statistic in the conditional power approach; $R^2_0$, "R squared" is the proportion of the variance in the dependent variable in the control arm that is predictable from $\bm{Z}=Z_1$; Power Loss, loss of power when conducting an interim analysis compared to the design power.
}
\end{figure}

First, the interim analysis is conducted at the fixed point in time when $75\%$ of the patients are expected to be recruited ($231$ patients in cohort $1$, $80$ patients in cohort $2$, $29$ patients in cohort $3$).
In Figure \ref{sim_CP_without}, the proposed conditional power approach incorporating baseline covariates is compared with standard conditional power for the different scenarios under true futility and true superiority, respectively. A comparison between both methods shows that incorporating baseline covariates leads to a \textit{free upgrade}: a higher information fraction as well as a higher probability to stop for true futility (Figure \ref{sim_CP_without_a}) are obtained with negligible loss of power (Figure \ref{sim_CP_without_b}).
The former is a consequence of the higher cutoff value corresponding with a higher information fraction.
Also note that, in general, a lower power (corresponding with higher $c$ and $R^2_0$) implies a higher probability to stop incorrectly for futility and a slightly higher power loss.
Comparing the different scenarios, shows that more predictive baseline covariates induce an increase in the information fraction and the probability to correctly stop for true futility. This is due to the efficiency gain from adjusting for baseline variables \cite{Qian2016}.

Second, an interim analysis is conducted when $50\%$\footnote{This information fraction is determined based on $Y$ only for the standard conditional power and based on $Z$, $X$ and $Y$ for the proposed method.} of the information in the simulated datasets is obtained in order to evaluate the time gain compared to the standard conditional power. Table \ref{sim2} shows that this information fraction is obtained earlier for the proposal, resulting in fewer recruited patients at the time of the interim analysis. Moreover, when using the proposed method, more predictive covariate(s) (higher $c$) lead to an earlier implementation of the interim analysis. The upper part of the table shows that the probability to stop incorrectly for futility as well as the loss of power is comparable between the two methods, while in the lower part it can be seen that the probability to stop correctly for true futility is similar. 
Thus, by using the proposal, fewer patients need to be recruited to make a decision with the same probability to correctly stop for true futility and with the same loss of power as with the standard method (power loss $<1\%$).
	\begin{table}[h!]
		\begin{threeparttable}
			\caption{\label{sim2}Comparison of the standard conditional power approach and the conditional power using the proposed interim estimator when the interim analysis is conducted at an information fraction of $50\%$. Upper table: results under superiority; Lower table: results under futility.}
			\centering
			\begin{tabular}{l l  ccccc}
				\hline	\hline	
				Model&Method && $\#$ Days & $\%$ Recruited & Prob. to Stop & Power Loss \\ \hline
				&&& &  &&\\
				\multirow{2}{*}{$c=0$}&Proposal  & &1288&74\%& 0.6\%&0.24\%\\
				&Standard CP & &1294&74\%& 0.5\%&0.16\%\\
				&&&  &&  &\\
				\multirow{2}{*}{$c=1$}&Proposal  & &1275&73\%& 0.8\%&0.48\%\\
				&Standard CP & &1294&74\%& 0.8\%&0.44\%\\
				&&&  &  &&\\
				\multirow{2}{*}{$c=2$}&Proposal  & &1243&72\%& 2.2\%&0.96\%\\
				&Standard CP & &1294&74\%& 2.5\%&0.88\%\\
				&&& & &&\\
				\multirow{2}{*}{$c=3$}&Proposal  & &1207&69\%& 5.4\%&0.92\%\\
				&Standard CP & &1294&74\%& 5.9\%&1.00\%\\
				&&&  &  &&\\
				\multirow{2}{*}{$c=4$}&Proposal  & &1173&68\%& 9.6\%&0.88\%\\
				&Standard CP & &1294&74\%& 9.7\%&0.92\%\\
				&&&   & &&\\
				\multirow{2}{*}{$c=5$}&Proposal  & &1144&66\%& 13.3\%&0.88\%\\
				&Standard CP & &1294&74\%& 13.6\%&0.92\%\\
				&  &  &&&&\\					
				\hline

				Model&Method && $\#$ Days & $\%$ Recruited & \multicolumn{2}{c}{Prob. to Stop } \\ \hline
				&&   &&&&\\			
\multirow{2}{*}{$c=0$}&Proposal  & &1289&74\%& \multicolumn{2}{c}{59.9\%}\\
&Standard CP & &1294&74\%& \multicolumn{2}{c}{60.4\%}\\
&&&  &  &&\\
\multirow{2}{*}{$c=1$}&Proposal  & &1279&74\%& \multicolumn{2}{c}{59.4\%}\\
&Standard CP & &1294&74\%& \multicolumn{2}{c}{59.2\%}\\
&&&  &  &&\\
\multirow{2}{*}{$c=2$}&Proposal  & &1253&72\%& \multicolumn{2}{c}{59.0\%}\\
&Standard CP & &1294&74\%& \multicolumn{2}{c}{59.6\%}\\
&&&  &  &&\\
\multirow{2}{*}{$c=3$}&Proposal  & &1222&70\%& \multicolumn{2}{c}{59.9\%}\\
&Standard CP & &1294&74\%& \multicolumn{2}{c}{60.1\%}\\
&&&  &  &&\\
\multirow{2}{*}{$c=4$}&Proposal  & &1191&69\%& \multicolumn{2}{c}{60.1\%}\\
&Standard CP & &1294&74\%& \multicolumn{2}{c}{60.0\%}\\
&&&  &  &&\\
\multirow{2}{*}{$c=5$}&Proposal  & &1164&67\%& \multicolumn{2}{c}{59.6\%}\\
&Standard CP & &1294&74\%& \multicolumn{2}{c}{60.0\%}\\
				&&&  &  &&\\
				\hline
			\end{tabular}
			\begin{tablenotes}
				\small
				\item Note: $\#$ Days, average number of days elapsed since beginning of the study; $\%$ Recruited, average percentage of patients already recruited; Prob. to Stop, probability to stop for futility using O'Brien Fleming boundaries; Power Loss, loss of power when conducting an interim analysis compared to the design power.
			\end{tablenotes}
		\end{threeparttable}
	\end{table}

\begin{figure}[h]
	\caption{Comparison of percentages of patients recruited at the time of the interim analysis between the standard conditional power and the porposed conditional power for different information fractions. Setting: Futility $c=2$.}
	\label{t_relativegain}
	\includegraphics[width=0.5\linewidth]{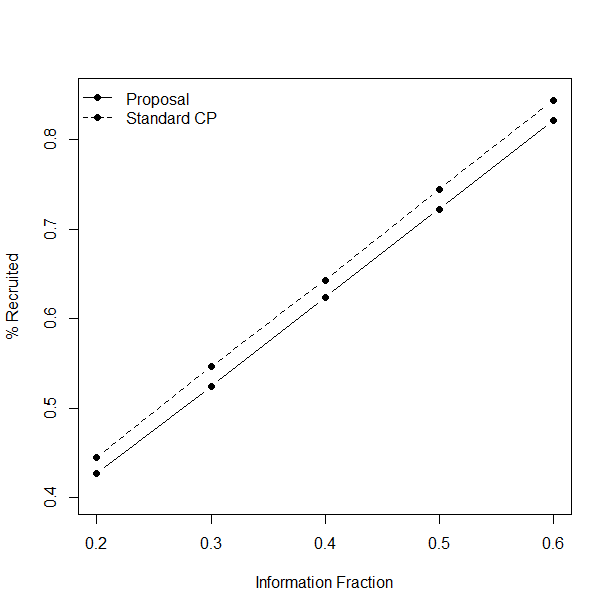}
\end{figure}

As can be seen in Figure \ref{t_relativegain}, conducting the interim analysis at different information fractions leads to similar absolute reductions in number of recruited patients. Therefore, the relative gain, defined as the difference in number of recruited patients between both methods relative to the number of recruited patients using the standard conditional power, increases with decreasing information fraction. 
Another determining factor is the average number of monthly recruited patients. We therefore simulated $2,500$ datasets under the futility scenario with $c=2$ for a recruitment rate of $16$ patients a month instead of $8$. The percentage of recruited patients at information fraction $0.30$ is $78\%$ and $73\%$ for the standard and proposed conditional power, respectively, corresponding with a relative gain of $6\%$. For a recruitment rate of $8$ patients a month, these equal $55\%$ and $52\%$, respectively, corresponding with a relative gain of $4\%$. The reduction in number of recruited patients is thus larger for faster recruitment settings. 

\subsubsection{With Short-Term Measurements}
To also incorporate short-term measurements along with three baseline covariates, the following data-generating mechanism $Y_i|X_i, \bm{Z}_i\overset{d}{=}Ber\{m_0(X_i, \bm{Z}_i)\}$ ($i = 1,\cdots , n$) was employed, with $m_0(X, \bm{Z})$ predictions from a logistic regression model for $Y$ involving the $2$-way interaction between $A$ and $X$, the $3$-way interaction between $A$, a continuous covariate $Z_1$ (same as in Section \ref{sim_without}) and a dichotomous covariate $Z_2$, the $3$-way interaction between $A$, a continuous covariate $Z_1$ (same as in Section \ref{sim_without}) and a $3$-level covariate $Z_3$, and all lower order terms.
Under this data-generating model, the marginal probabilities of success in the control and experimental treatment arm are $0.60$ and $0.49$, respectively. To attain $90\%$ power at a one-sided significance level of $2.5\%$, $421$ patients in both arms are required.

Simulation experiments with a correctly specified outcome model used the working models $h_1(X, \bm{Z}, \bm{\eta})=\text{logit}^{-1}(\eta_1+\eta_2X+\eta_3Z_1 +\eta_4Z_2+\eta_5Z_3+\eta_6Z_1Z_2+\eta_7Z_1Z_3)$ and $h_0(X, \bm{Z}, \bm{\zeta})=\text{logit}^{-1}(\zeta_1+\zeta_2X+\zeta_3Z_1 +\zeta_4Z_2+\zeta_5Z_3+\zeta_6Z_1Z_2+\zeta_7Z_1Z_3)$ to predict the missing outcomes in cohort $2$ and working models $f_1(\bm{Z}, \bm{\delta})=\text{logit}^{-1}(\delta_1+\delta_2Z_1 +\delta_3Z_2+\delta_4Z_3+\delta_5Z_1Z_2+\delta_6Z_1Z_3)$ and $f_0(\bm{Z}, \bm{\nu})=\text{logit}^{-1}(\nu_1+\nu_2Z_1 +\nu_3Z_2+\nu_4Z_3+\nu_5Z_1Z_2+\nu_6Z_1Z_3)$ to predict the missing outcomes in cohort $3$.
To evaluate the performance when the prediction models are misspecified and to investigate whether this could lead to incorrectly stopping, we also considered the following outcome working models
\begin{enumerate}
	\item a misspecified model including $X$, $Z_1$, $Z_2$ and $Z_3$ but without interactions; $h_1(X, \bm{Z}, \bm{\eta})=\text{logit}^{-1}(\eta_1+\eta_2X+\eta_3Z_1 +\eta_4Z_2+\eta_5Z_3)$ and $h_0(X, \bm{Z}, \bm{\zeta})=\text{logit}^{-1}(\zeta_1+\zeta_2X+\zeta_3Z_1 +\zeta_4Z_2+\zeta_5Z_3)$ to predict the missing outcomes in cohort $2$ and $f_1(\bm{Z}, \bm{\delta})=\text{logit}^{-1}(\delta_1+\delta_2Z_1 +\delta_3Z_2+\delta_4Z_3)$ and $f_0(\bm{Z}, \bm{\nu})=\text{logit}^{-1}(\nu_1+\nu_2Z_1 +\nu_3Z_2+\nu_4Z_3)$ to predict the missing outcomes in cohort $3$;
	\item a model only including $X$ and $Z_1$; $h_1(X, \bm{Z}, \bm{\eta})=\text{logit}^{-1}(\eta_1+\eta_2X+\eta_3Z_1)$ and $h_0(X, \bm{Z}, \bm{\zeta})=\text{logit}^{-1}(\zeta_1+\zeta_2X+\zeta_3Z_1)$ to predict the missing outcomes in cohort $2$ and $f_1(\bm{Z}, \bm{\delta})=\text{logit}^{-1}(\delta_1+\delta_2Z_1)$ and $f_0(\bm{Z}, \bm{\nu})=\text{logit}^{-1}(\nu_1+\nu_2Z_1)$ to predict the missing outcomes in cohort $3$;
	\item a misspecified model including $X$ and the absolute value of $Z_1$; $h_1(X, \bm{Z}, \bm{\eta})=\text{logit}^{-1}(\eta_1+\eta_2X+\eta_3|Z_1|)$ and $h_0(X, \bm{Z}, \bm{\zeta})=\text{logit}^{-1}(\zeta_1+\zeta_2X+\zeta_3|Z_1|)$ to predict the missing outcomes in cohort $2$ and $f_1(\bm{Z}, \bm{\delta})=\text{logit}^{-1}(\delta_1+\delta_2|Z_1|)$ and $f_0(\bm{Z}, \bm{\nu})=\text{logit}^{-1}(\nu_1+\nu_2|Z_1|)$ to predict the missing outcomes in cohort $3$;
	\item a model only including $X$ and $Z_3$; $h_1(X, \bm{Z}, \bm{\eta})=\text{logit}^{-1}(\eta_1+\eta_2X+\eta_3Z_3)$ and $h_0(X, \bm{Z}, \bm{\zeta})=\text{logit}^{-1}(\zeta_1+\zeta_2X+\zeta_3Z_3)$ to predict the missing outcomes in cohort $2$ and $f_1(\bm{Z}, \bm{\delta})=\text{logit}^{-1}(\delta_1+\delta_2Z_3)$ and $f_0(\bm{Z}, \bm{\nu})=\text{logit}^{-1}(\nu_1+\nu_2Z_3)$ to predict the missing outcomes in cohort $3$.
\end{enumerate}
\begin{figure}[h!]
	\centering
	\begin{subfigure}{.5\textwidth}
		\centering
		\includegraphics[width=.9\linewidth]{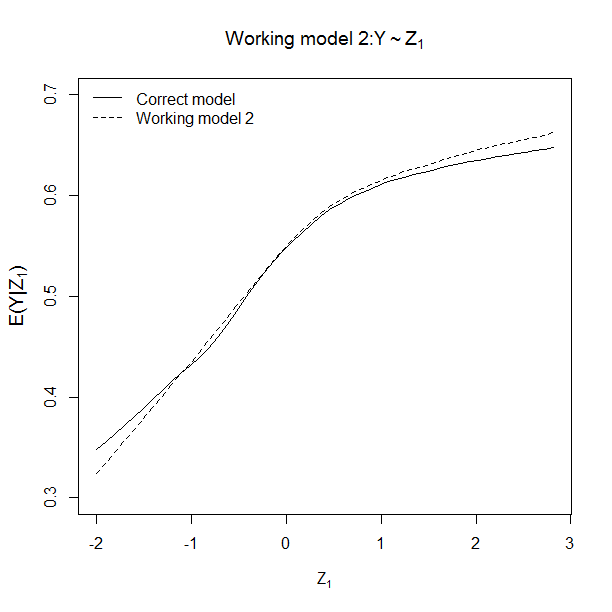}
		\caption{Outcome working model $2$}
	\end{subfigure}%
	\begin{subfigure}{.5\textwidth}
		\centering
		\includegraphics[width=.9\linewidth]{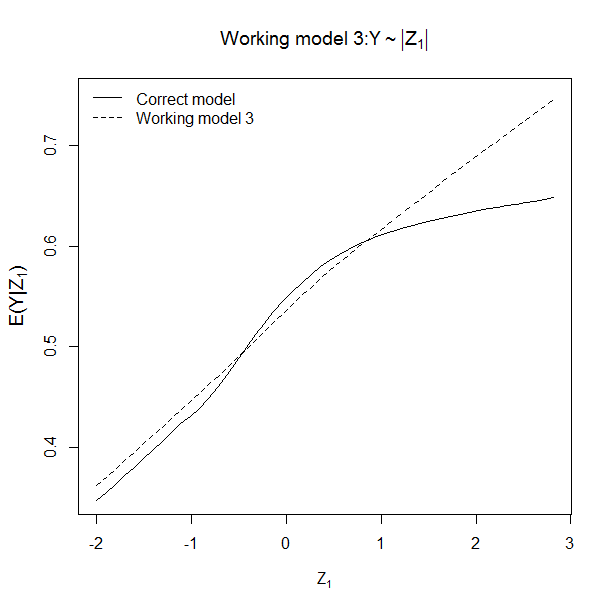}
		\caption{Outcome working model $3$}
	\end{subfigure}
	\caption{Visualisation of the model misspecfication for outcome working models $2$ and $3$: expected $Y$ values under the correctly specified model and outcome working models $2$ and $3$ are plotted against $Z_1$.}
	\label{Misspecification_visual}
\end{figure}
The expected $Y$ values for outcome working models $2$ and $3$ as well as for the correctly specified model are plotted against $Z_1$ in Figure \ref{Misspecification_visual}.
To compare our proposal to methods only using $X$ (e.g.\,Niewczas et al.\,, $2016$), outcome working models only including $X$ were used. Note that, in that case, we don't predict the outcome for the patients in cohort $3$.
The proportion of the variance in $Y$ that is related to $\bm{Z}=\{Z_1, Z_2, Z_3\}$, $X$ and both under these different prediction models is shown in Table \ref{R2_cp2}. 

	\begin{table}
	\begin{threeparttable}
		\caption{\label{R2_cp2} 
			Empirical $R$-squared values for the data generated under the settings for the conditonal power with short-term endpoint measurements. Simulations results are based on $10,000$ Monte Carlo replications.
		}
		\centering
			\begin{tabular}{l l  ccc c ccc}
				\hline	\hline	
				&&\multicolumn{7}{c}{$A=1$}\\
				\cline{3-9} \cline{7-9}\\
				&&\multicolumn{3}{c}{$H_0$}&&\multicolumn{3}{c}{$H_1$}\\		
				\cline{3-5}  \cline{7-9}\\
				Model&& $R^2_{Z}$ & $R^2_{X}$ & $R^2$& & $R^2_{Z}$ & $R^2_{X}$ & $R^2$\\ \hline
				&&& &&  \\
				Correct&& $0.27$ & $0.34$ & $0.61$& & $0.28$ & $0.72$ & $1$\\
				Misspecified ($1$)&& $0.11$ & $0.40$ & $0.52$& & $0.09$ & $0.66$ & $0.74$\\
				Misspecified ($2$)&& $0.02$ & $0.40$ & $0.42$& & $0.06$ & $0.66$ & $0.72$\\
				Misspecified ($3$)&& $0.004$ & $0.39$ & $0.40$& & $0.004$ & $0.71$ & $0.72$ \\
				Misspecified ($4$)&& $0.10$ & $0.39$ & $0.48$& & $0.01$ & $0.71$ & $0.72$\\
				Only $X$&&$0$ & $0.39$ & $0.39$& & $0$ & $0.71$ & $0.71$\\		
				&&& &&  \\ \hline
			\end{tabular}
		\begin{tablenotes}
			\small
			\item 
			Note: $R^2_Z$, $R^2_X$ and $R^2$ correspond to the proportion of the variance in $Y$ that is predictable from respectively $\bm{Z}=\{Z_1, Z_2, Z_3\}$, $X$ and both $\bm{Z}$ and $X$; estimated using formulas in Appendix E.
		\end{tablenotes}
	\end{threeparttable}
\end{table}

First, the interim analysis is conducted at the fixed point in time when $75\%$ of the patients are expected to be recruited ($407$ patients in cohort $1$, $164$ patients in cohort $2$, $60$ patients in cohort $3$). 
Table \ref{cp_with_fixed} shows that a higher total $R^2$ (see Table \ref{R2_cp2}) and the use of correctly specified working models, lead to a higher information fraction and a higher probability to correctly stop for true futility. This is again due to the efficiency gain from adjusting for (more predictive) baseline variables and short-term endpoints (\cite{Qian2016}).

When conducting an interim analysis at the time $50\%$ of the information is obtained, the advantage over the standard conditional power increases with increasing total $R^2$, while the advantage over the method only using $X$ is mainly determined by $R^2_Z$. The latter is a consequence of the fact that the proposal also incorporates baseline covariates on top of the short-term endpoint.
However, in this simulation study there are on average only $60$ patients in cohort $3$ compared to $164$ in cohort $2$ and $387$ in cohort $1$. Therefore, the relative advantage over the method only using $X$ compared to the advantage of the latter over the standard method, is rather limited. The relative gain will increase when cohort $3$ contains proportionately more patients.
To see this, we conducted a simulation study where the short-term endpoint is measured $12$ months after randomization instead of $4$ months, corresponding with an average of $404$ patients in cohort $1$, $75$ in cohort $2$ and $149$ in cohort $3$. The percentage of recruited patients is then given by $77\%$, $75\%$ and $72\%$ for respectively the standard conditional power, the conditional power only including the short-term endpoint $X$ and our proposal. The absolute advantage ($3\%$) of including both baseline covariates and the short-term endpoint over the method only incorporating the short-term endpoint is similar as for $X$ measured $4$ months after randomization. The gain in terms of percentage recruited patients at the time of the interim analysis by incorporating baseline covariates on top of the short-term endpoint relative to the gain by only incorporating the short-term endpoint increases from $1.6$ $(=(77-69)/(77-72))$ to $2.5$ $(=(77-72)/(77-75))$, meaning that the decrease in number of recruited patients compared to the standard conditional power is respectively $1.6$ and $2.5$ times larger for the proposal than for the conditional power only including $X$ and $Y$.


\begin{table}[h!]
	\begin{threeparttable}
		\caption{\label{cp_with_fixed}Comparison when the interim analysis is conducted at the fixed point in time where $75\%$ of the patients are recruited. Upper table: results under superiority; Lower table: results under futility.}
		\centering
		\begin{tabular}{ lcccc}
			\hline	\hline	
			Method && $t$ & Prob. to Stop & Power Loss\\ \hline
			&&\multicolumn{3}{c}{$\#$Cohort $1=407$, $\#$Cohort $2=164$, $\#$Cohort $3=60$}\\ \cline{3-5}
			&&  &  &\\
			Proposal, correct  & &$0.58$&$3.3\%$& $0.84\%$\\
			Proposal, misspecified ($1$)  & &$0.56$&$3.0\%$& $0.92\%$\\
			Proposal, misspecified ($2$)  & &$0.54$&$2.8\%$& $0.76\%$\\
			Proposal, misspecified ($3$)  & &$0.54$&$3.0\%$& $0.80\%$\\
			Proposal, misspecified ($4$)  & &$0.55$&$2.9\%$& $0.80\%$\\
			Proposal, only $X$ & &$0.54$&$3.0\%$& $0.80\%$\\
			Standard CP, only $Y$ & &$0.48$&$1.8\%$& $0.48\%$\\
			&&  &  &\\
			\hline
			Method && $t$ &\multicolumn{2}{c}{Prob. to Stop}\\ \hline
			&&\multicolumn{3}{c}{$\#$Cohort $1=407$, $\#$Cohort $2=164$, $\#$Cohort $3=60$}\\ \cline{3-5}
			&&  &  &\\
			Proposal, correct  & & $0.56$ &\multicolumn{2}{c}{$69.4\%$} \\
			Proposal, misspecified ($1$)  & &$0.55$&\multicolumn{2}{c}{$68.0\%$} \\
			Proposal, misspecified ($2$)  & &$0.53$&\multicolumn{2}{c}{$65.1\%$} \\
			Proposal, misspecified ($3$)  & &$0.53$&\multicolumn{2}{c}{$64.8\%$} \\
			Proposal, misspecified ($4$)  & &$0.54$&\multicolumn{2}{c}{$67.0\%$} \\
			Proposal, only $X$ & &$0.53$&\multicolumn{2}{c}{$64.4\%$} \\
			Standard CP, only $Y$ & & $0.48$ &\multicolumn{2}{c}{$56.4\%$} \\
			&&&&\\
			\hline
		\end{tabular}
		\begin{tablenotes}
			\small
			\item Note: $t$, average information fraction; Prob. to Stop, probability to stop for futility using O'Brien Fleming boundaries; Power Loss, loss of power when conducting an interim analysis compared to the design power.
		\end{tablenotes}
	\end{threeparttable}
\end{table}

	\begin{table}[h!]
	\begin{threeparttable}
		\caption{\label{cp_with_monit}Comparison when the interim analysis is conducted at an information fraction of $50\%$. Upper table: results under superiority; Lower table: results under futility.}
		\centering
		\begin{tabular}{ lccccc}
			\hline	\hline	
			Method &&$\#$ Days & $\%$ Recruited & Prob. to Stop & Power Loss \\ \hline
			&&  &  &&\\
			Proposal, correct  & &$1066$&$67\%$& $2.3\%$&$0.44\%$\\
			Proposal, misspecified ($1$)  & &$1101$&$69\%$&$ 2.4\%$&$0.60\%$\\
			Proposal, misspecified ($2$)  & &$1119$&$71\%$& $2.2\%$&$0.52\%$\\
			Proposal, misspecified ($3$)  & &$1124$&$71\%$& $2.1\%$&$0.52\%$\\
			Proposal, misspecified ($4$)  & &$1111$&$70\%$& $2.2\%$&$0.52\%$\\
			Proposal, only $X$  & &$1126$&$71\%$& $2.1\%$&$0.56\%$\\
			Standard CP, only $Y$  & &$1217$&$77\%$& $2.0\%$&$0.48\%$\\	
						&&  &  &&\\				
			\hline
						Method &&$\#$ Days & $\%$ Recruited & \multicolumn{2}{c}{Prob. to Stop } \\ \hline
			&&  &  &&\\
			Proposal, correct   & &$1096$&$69\%$& \multicolumn{2}{c}{$60.3\%$}\\
			Proposal, misspecified ($1$)   & &$1116$&$70\%$& \multicolumn{2}{c}{$60.0\%$}\\
			Proposal, misspecified ($2$)   & &$1142$&$72\%$& \multicolumn{2}{c}{$59.1\%$}\\
			Proposal, misspecified ($3$)   & &$1145$&$72\%$& \multicolumn{2}{c}{$59.9\%$}\\
			Proposal, misspecified ($4$)   & &$1124$&$71\%$& \multicolumn{2}{c}{$59.2\%$}\\
			Proposal, only $X$   & &$1146$&$72\%$& \multicolumn{2}{c}{$59.7\%$}\\
			Standard CP, only $Y$  & &$1217$&$77\%$& \multicolumn{2}{c}{$59.3\%$}\\
			&&  &&&\\
			\hline
		\end{tabular}
		\begin{tablenotes}
			\small
			\item Note: $\#$ Days, average number of days elapsed since beginning of the study; $\%$ Recruited, average percentage of patients already recruited; Prob. to Stop, probability to stop for futility using O'Brien Fleming boundaries; Power Loss, loss of power when conducting an interim analysis compared to the design power.
		\end{tablenotes}
	\end{threeparttable}
\end{table}



\newpage
\subsection{Simulations on Sample Size Re-Estimation}
In these simulation studies, futility stopping based on conditional power as well as sample size adaptations were performed. 
The total sample size was bounded to be at most $2$ times the original sample size.
We generated data under a data-generating mechanism where for each $i$ ($i = 1,\cdots , n$), $Y_i|X_i, A_i, \bm{Z}	_i\overset{d}{=}Ber(m_0(X_i, A_i, \bm{Z}_i))$, with  $m_0(X, A, \bm{Z})=\text{logit}^{-1}(\omega_0+c\cdot\omega_1A+\omega_2X+c\cdot\omega_3XA+\omega_4Z_1+c\cdot\omega_5Z_1A)$, $c\in\{0, 0.5, 1,5, 1.5\}$ and the coeffcients $\{\omega_0, \dots, \omega_5\}$ correspond to these obtained from a logistic regression fitted on the original phase $2$b data. 
The marginal probabilities of success in the treatment and control arm under the data-generating model with $c=1$ are $0.60$ and $0.47$, respectively, corresponding with a treatment effect of $0.13$. To attain a power of $0.90$ under this alternative at a one-sided significance level of $0.025$ the required sample size in both treatment arms is $319$ patients. The other probabilities of success in the treatment arm are $0.47$, $0.56$ and $0.63$ for $c=0$, $c=0.5$ and $c=1.5$, respectively, corresponding with a treatment effect of respectively $0$, $0.09$ and $0.16$, respectively.

Table \ref{ssr_fixed_obs} shows the results for a simulation study where the interim analysis is conducted at the fixed point in time where $478$ out of the planned $638$ patients, corresponding with $75\%$, are recruited.  We can see that in all cases the type I error rate is controlled, since the power after sample size reassessment is around $2.5\%$ for a treatment effect equal to $0$. As before, the information fraction as well as the probability to stop for true futility are higher for the proposal. However, this also causes a slightly larger loss in power relative to the standard approach. When sample size re-assessment is performed, there is an increase in power for both methods compared to the designs without sample size reassessment and the classical one-stage trial, except for a treatment effect equal to $0.16$.  
On average a smaller sample size is needed for the proposal, but this comes with a lower overall power over the two stages. The lower sample size causes a higher probability to be unsuccessful after sample size reassessment for a trial that is not stopped for futility and successful without sample size reassessment, which results in a lower power.

The results for the simulation study where the interim analysis is conducted when $50\%$ of the total information is obtained, are summarised in Table \ref{ssr_monit_obs} in Appendix F. The probability to stop for futility and the power loss when no sample size reassessment is performed are more comparable in this simulation study, but generally the results are similar for the setting where the interim analysis is conducted at a fixed point in time.

Doing a sample size reassessment after performing an interim analysis thus seems to be equally beneficial for both methods; resulting in (slightly) lower sample sizes at a cost of (slightly) lower overall power for the proposal compared to the standard method. Given the lower total sample size and lower number of patients recruited at the time of the interim analysis compared to the standard method, it seems advantageous to use the proposal, even with the cost of a (slightly) lower power.


	\begin{table}
	\begin{threeparttable}
		\caption{\label{ssr_fixed_obs}Operating characteristics of a trial with sample size re-assessment based on conditional power as futility stopping rule at the fixed point in time where $478$ out of the planned $638$ patients are recruited. Results are based on $5000$ Monte Carlo simulations.}
		\centering
		\begin{tabular}{l l c c c c }
			\hline	\hline
			&&\multicolumn{4}{c}{Treatment Effect}\\	
			&&$0$& $0.09$  & $0.13$ & $0.16$\\ 
			\hline
			&Power One Stage Trial&$2.52\%$&$57.78\%$  &$90.66\%$&$98.70\%$\\ \hline
			\textbf{Proposal}&Average $t$&$0.46$&$0.47$ &$0.48$&$0.49$\\
			&Probability to Stop for Futility&$50.0\%$&$8.0\%$ &$1.9\%$&$0.4\%$\\
			&Power Loss&$ 0.04\%$& $0.65\%$&$0.76\%$&$0.19\%$\\
			&Power No SSR&$2.47\%$& $57.13\%$ &$89.90\%$&$98.51\%$\\
			&$Q_0$ No SSR&$478$& $478$ &$478$&$478$\\ 
			&$Q_1$ No SSR&$478$& $638$ &$638$&$638$\\
			&$Q_2$ No SSR&$478$& $638$ &$638$&$638$\\
			&$Q_3$ No SSR&$638$& $638$ &$638$&$638$\\
			&$Q_4$ No SSR&$638$& $638$ &$638$&$638$\\
			&ASS No SSR&$557$($80$)& $625$($43$) &$635$($22$)&$637$($9$)\\
			\cline{2-6}
			&Power SSR&$2.51\%$& $73.75\%$ &$93.96\%$&$98.62\%$\\
			&$Q_0$ SSR&$478$&$478$  &$478$&$478$\\ 
			&$Q_1$ SSR&$478$& $478$ &$478$&$478$\\
			&$Q_2$ SSR&$478$& $872$ &$484$&$478$\\
			&$Q_3$ SSR&$1276$& $1276$ &$964$&$510$\\
			&$Q_4$ SSR&$1276$& $1276$ &$1276$&$1276$\\
			&ASS SSR&$837$($389$)& $890$($359$) &$715$($320$)&$581$($230$)\\
			&\% Rejected with SSR, not without&$2.7\%$& $59.6\%$ &$82.8\%$&$93.0\%$\\
			&\% Not rejected with SSR, rejected without&$50.0\%$& $7.3\%$ &$3.0\%$&$1.0\%$\\
			\hline
			\textbf{Standard CP}&Average $t$&$0.41$&$0.41$ &$0.41$&$0.41$\\
			&Probability to Stop for Futility&$38.6\%$&$4.87\%$ &$1.13\%$&$0.19\%$\\
			&Power Loss&$0.04\%$& $0.45\%$&$0.42\%$&$0.09\%$\\
			&Power No SSR&$2.50\%$& $57.33\%$ &$90.24\%$&$98.61\%$\\
			&$Q_0$ No SSR&$478$& $478$ &$478$&$478$\\ 
			&$Q_1$ No SSR&$478$& $638$ &$638$&$638$\\
			&$Q_2$ No SSR&$638$& $638$ &$638$&$638$\\
			&$Q_3$ No SSR&$638$& $638$ &$638$&$638$\\
			&$Q_4$ No SSR&$638$& $638$ &$638$&$638$\\
			&ASS No SSR&$576$($78$)& $630$($34$) &$636$($17$)&$638$($7$)\\
			\cline{2-6} 
			&Power SSR&$2.63\%$& $75.12\%$ &$95.65\%$&$99.07\%$\\
			&$Q_0$ SSR&$478$&$478$  &$478$&$478$\\ 
			&$Q_1$ SSR&$478$& $478$ &$478$&$478$\\
			&$Q_2$ SSR&$1276$& $971$ &$508$&$478$\\
			&$Q_3$ SSR&$1276$& $1276$ &$1145$&$554$\\
			&$Q_4$ SSR&$1276$& $1276$ &$1276$&$1211$\\
			&ASS SSR&$924$($386$)& $917$($357$) &$742$($335$)&$605$($256$)\\
			&\% Rejected with SSR, not without&$2.2\%$& $56.0\%$ &$83.3\%$&$88.3\%$\\
			&\% Not rejected with SSR, rejected without&$46.4\%$& $5.9\%$ &$2.0\%$&$0.6\%$\\  \hline				
			\hline
			&&  &&&
		\end{tabular}
		\begin{tablenotes}
			\small
			\item Note: Power One Stage Trial, a reference power of one stage trial where no interim analyses or adaptations are conducted; Average $t$, average time fraction; FS, probability to stop for futility using O'Brien Fleming boundaries; Power Loss, loss of power when conducting an interim analysis compared to the design power; Power SSR/No SSR, Power when there is/is no sample size re-assessment; $Q_i$ SSR/No SSR, $i^{th}$ quantile of the sample size when there is/is no sample size re-assessment; ASS SSR/No SSR, average sample size over both stages and its standard deviation (in brackets); $\%$ Rejected with SSR, not without, proportion of trials that were not stopped and not rejected without SSR that are rejected with SSR; $\%$ Not rejected with SSR, rejected without, proportion of trials that were not stopped and rejected without SSR that are not rejected with SSR.	
		\end{tablenotes}
	\end{threeparttable}
\end{table}


\section{Discussion}
In this paper, we propose a method to efficiently exploit all information available on baseline covariates as well as short-term endpoint data (e.g.\, an assessment of the primary endpoint at an earlier time) 
to improve decision-making as well as sample size reassessment at an interim analysis. 
Our proposal assumes that the observed association between the primary endpoint and the short-term endpoint and the observed association between the primary endpoint and the baseline covariates 
do not change over time. The plausibility of this assumption, which can often be expected to hold under random recruitment, should be assessed a priori based on consultation with clinicians.
In contrast to competing approaches using surrogate endpoints (e.g.\,, Prentice, $1989$), the estimand that we consider at the interim analysis is the treatment effect on the final endpoint, rather than on the short-term endpoint. This has the advantage that no assumptions are needed on the extent to which the short-term endpoint is a good surrogate for the final endpoint. It moreover enables one to better judge the clinical significance of the interim results. It has the drawback, however, that data will be needed on patients who have completed the study, even at the interim assessment. 

Our proposed estimator has the appealing feature that model misspecification does not introduce bias, but it may reduce efficiency compared to a correctly specified prediction model. 
Despite this precision loss, our proposed estimator is never outperformed by the standard analyses that only use information on the short-term and primary endpoint since the proposed estimator reduces to these standard analyses if no baseline covariates are available.
Moreover, when the models are correct, the resulting estimator is asymptotically efficient in the subclass of estimators that have this robustness.
Implementing the interim test statistic based on the proposed estimator in the conditional power approach 
therefore allows earlier stopping for true futility whilst controlling the probability for incorrectly stopping, resulting in fewer recruited patients. 
Additionally, sample size reassessment based on conditional power was considered. To guarantee the type I error, 
the combination test was applied for the final analysis.
In simulation studies, we observed that the incorporation of information on baseline covariates and short-term endpoints is beneficial in terms of smaller sample sizes needed to obtain a similar power.
A potential concern was that the additional information accrued between 
the interim analysis and the final analysis potentially involves observations 
of subjects who already contributed information to the interim analysis. 
Building on Scharfstein et al.\,($1997$), we have shown that this is not problematic for the considered test statistic, as it satisfies the independent increments property.

Although the present paper has focused on binary endpoints, the same approach can be used for continuous and survival endpoints. 
In future work, we will use repeated measures models that use information at multiple secondary endpoint times to provide furter efficiency gains.

At early interim analyses, incorporating the most predictive baseline covariate(s) or short term endpoint(s) may already lead to an important efficiency gain. However, we recommend using increasingly flexible models as more primary endpoint data come available. This because an asymptotic efficiency gain is guaranteed upon adding more terms to the model (and thus increasing the $R^2$) even when the model is not correctly specified.  

For simplicity, we have developed our proposal so that the interim analysis coincides with the standard analysis at the end. 
However, there is still a high potential to gain efficiency, and thus reduce the required sample size for the study, by also including baseline covariates in the final analysis.  
This can be done by using efficient estimators similar to those discussed in this paper for the end of study comparison (Tsiatis, $2008$).

\bibliography{biblio_correct}{}

\begin{thebibliography}{10}

\bibitem{Assmann2000}
{\sc Assmann, S.~F., Pocock, S.~J., Enos, L.~E., and Kasten, L.~E.}
\newblock Subgroup analysis and other (mis)uses of baseline data in clinical
  trials.
\newblock {\em The Lancet 355}, 9209 (2000), 1064 -- 1069.

\bibitem{Austin2010}
{\sc Austin, P.~C., Manca, A., Zwarenstein, M., Juurlink, D.~N., and Stanbrook,
  M.~B.}
\newblock A substantial and confusing variation exists in handling of baseline
  covariates in randomized controlled trials: a review of trials published in
  leading medical journals.
\newblock {\em Journal of Clinical Epidemiology 63}, 2 (2010), 142 -- 153.

\bibitem{Bauer1989}
{\sc Bauer, P.}
\newblock Multistage testing with adaptive designs.
\newblock {\em Biometrie und Informatik in Medizin und Biologie 20\/} (1989),
  130–148.

\bibitem{Bauer1994}
{\sc Bauer, P., and K\"ohne, K.}
\newblock Evaluation of experiments with adaptive interim analyses.
\newblock {\em Biometrics 50}, 4 (1994), 1029--1041.

\bibitem{Bauer2006}
{\sc Bauer, P., and K\"onig, F.}
\newblock The reassessment of trial perspectives from interim data - a critical
  review.
\newblock {\em Statistics in Medicine 25\/} (2006), 23–36.

\bibitem{Bretz2009}
{\sc Bretz, F., K\"onig, F., Brannath, W., Glimm, E., and Posch, M.}
\newblock Tutorial in biostatistics: Adaptive designs for confirmatory clinical
  trials.
\newblock {\em Statistics in Medicine 28\/} (2009), 1181–1217.

\bibitem{Diacon}
{\sc Diacon, A.~H., Pym, A., Grobusch, M.~P., and et~al}.
\newblock Multidrug-resistant tuberculosis and culture conversion with
  bedaquiline.
\newblock {\em New England Journal of Medicine 371}, 8 (2014), 723--732.
\newblock PMID: 25140958.

\bibitem{Halperin1982}
{\sc Halperin, M., Lan, K. K.~G., Ware, J.~H., Johnson, N.~J., and DeMets,
  D.~L.}
\newblock An aid to data monitoring in long-term clinical trials.
\newblock {\em Controlled Clinical Trials 3}, 4 (1982), 311 -- 323.

\bibitem{HampsonJennison2013}
{\sc Hampson, L.~V., and Jennison, C.}
\newblock Group sequential tests for delayed responses (with discussion).
\newblock {\em Journal of the Royal Statistical Society: Series B (Statistical
  Methodology) 75}, 1 (2013), 3--54.

\bibitem{Lachin2005}
{\sc Lachin, J.~M.}
\newblock A review of methods for futility stopping based on conditional power.
\newblock {\em Statistics in Medicine 24}, 18 (2005), 2747--2764.

\bibitem{LanWittes1988}
{\sc Lan, K. K.~G., and Wittes, J.}
\newblock The b-value: A tool for monitoring data.
\newblock {\em Biometrics 44}, 2 (1988), 579--585.

\bibitem{Lehmacher1999}
{\sc Lehmacher, W., and Wassmer, G.}
\newblock Adaptive sample size calculations in group sequential trials.
\newblock {\em Biometrics 55}, 4 (1999), 1286--1290.

\bibitem{Moore2009}
{\sc Moore, K.~L., and van~der Laan, M.~J.}
\newblock Covariate adjustment in randomized trials with binary outcomes:
  targeted maximum likelihood estimation.
\newblock {\em Statistics in medicine 28 1\/} (2009), 39--64.

\bibitem{Niewczas2016}
{\sc Niewczas, J., Kunz, C.~U., and K\"onig, F.}
\newblock Interim analysis incorporating short- and long-term binary endpoints.
\newblock {\em Biometrical Journal\/} (2019), 1--23.

\bibitem{Pocock2002}
{\sc Pocock, S., E~Assmann, S., E~Enos, L., and Kasten, L.}
\newblock Subgroup analysis, covariate adjustment and baseline comparisons in
  clinical trial reporting: Current practice and problems.
\newblock {\em Statistics in medicine 21}, 19 (2002), 2917--30.

\bibitem{Prentice1989}
{\sc Prentice, R.~L.}
\newblock Surrogate endpoints in clinical trials: Definition and operational
  criteria.
\newblock {\em Statistics in Medicine 8}, 4 (1989), 431--440.

\bibitem{proschan2006}
{\sc Proschan, M., Lan, K., and Wittes, J.}
\newblock {\em Statistical Monitoring of Clinical Trials: A Unified Approach}.
\newblock Statistics for Biology and Health. Springer New York, 2006.

\bibitem{Qian2016}
{\sc Qian, T., and Rosenblum, M.}
\newblock Improving power in group sequential, randomized trials by adjusting
  for prognostic baseline variables and short-term outcomes.
\newblock {\em Johns Hopkins University, Dept. of Biostatistics Working Papers.
  Working Paper 285. 5}, 5 (2016), 5.

\bibitem{Scharfstein1997}
{\sc Scharfstein, D.~O., Tsiatis, A.~A., and Robins, J.~M.}
\newblock Semiparametric efficiency and its implication on the design and
  analysis of group-sequential studies.
\newblock {\em Journal of the American Statistical Association 92}, 440 (1997),
  1342--1350.

\bibitem{Stallard2010}
{\sc Stallard, N.}
\newblock A confirmatory seamless phase ii/iii clinical trial design
  incorporating short-term endpoint information.
\newblock {\em Statistics in Medicine 29}, 9 (2010), 959--971.

\bibitem{Tsiatis2006}
{\sc Tsiatis, A.~A.}
\newblock {\em Semiparametric theory and missing data}.
\newblock Springer, New York, 2006.

\bibitem{Tsiatis2008}
{\sc Tsiatis, A.~A., Davidian, M., Zhang, M., and Lu, X.}
\newblock Covariate adjustment for two-sample treatment comparisons in
  randomized clinical trials: a principled yet flexible approach.
\newblock {\em Statistics in medicine 27 23\/} (2008), 4658--77.

\bibitem{wassmer2016}
{\sc Wassmer, G., and Brannath, W.}
\newblock {\em Group Sequential and Confirmatory Adaptive Designs in Clinical
  Trials}.
\newblock Springer Series in Pharmaceutical Statistics. Springer International
  Publishing, 2016.

\bibitem{Yang2001}
{\sc Yang, L., and A~Tsiatis, A.}
\newblock Efficiency study of estimators for a treatment effect in a
  pretest-posttest trial.
\newblock {\em The American Statistician 55\/} (02 2001), 314--321.

\bibitem{Zhang2010}
{\sc Zhang, Y., and Clarke, W.~R.}
\newblock A flexible futility monitoring method with time-varying conditional
  power boundary.
\newblock {\em Clinical Trials 7}, 3 (06 2010), 209--18.

\end{thebibliography}

\newpage
\section*{Appendix}
\subsection*{Appendix A: The Interim Estimator}
The focus in this section is on the estimator $\hat{\mu}_1$ in the experimental treatment arm, but reasonings and results are similar for $\hat{\mu}_0$ and $\hat{\mu}_1-\hat{\mu}_0$. The formulas in this section make use of the estimated unadjusted probabilities $\hat{\pi}$, $\hat{\pi}^X$ and $\hat{\pi}^Y$.  
\subsubsection*{Appendix A.1: Theoretical Derivation of the Interim Estimator}
Let $\mu_1=E(Y|A=1)$ and consider the problem of estimating $\mu_1$ using the observed data. Under the non-parametric model for the data ($A$, $Y$, $X$, $\bm{Z}$), all RAL estimators of $\mu_1$ have influence function equal to
\begin{align*}
\varphi^{(0)}=(A/\pi)(Y-\mu_1),
\end{align*}
with $\pi$ the true randomization probability. 
Assume that the covariates $\bm{Z}$ and the treatment indicator $A$ contain sufficient information to explain the missingness so that the missing at random (MAR) assumption, $Y\indep C^X|A, \bm{Z}$ (Tsiatis, 2006), holds. Tsiatis (2006) shows that all RAL estimators of $\mu_1$, under the assumption that $\pi^X$ is known, have influence function
\begin{align}\label{phis2}
\left\{\frac{C^X\varphi^{(0)}}{\pi^X}+\left(1-\frac{C^X}{\pi^X}\right)L^{(1)}(A, \bm{Z})\right\},
\end{align}
where $L^{(1)}(A, \bm{Z})$ is an arbitrary function of $A$ and $\bm{Z}$. 
From Theorem $10.2$ in Tsiatis (2006), for given $\varphi^{(0)}$, the optimal choice for $L^{(1)}(A, \bm{Z})$ that leads to a semi-parametric efficient estimator under the above restrictions is given by $E(\varphi^{(0)}|C^X=1, A, \bm{Z})$.
Therefore, among the class of influence functions (\ref{phis2}), the one with the smallest variance is
\begin{align}\label{phi2}
\begin{split}
\varphi^{(1)} &= \frac{C^X\varphi^{(0)}}{\pi^X}+\left(1-\frac{C^X}{\pi^X}\right)E(\varphi^{(0)}|C^X=1, A, \bm{Z})\\
&= \frac{C^XA}{\pi^X\pi}Y+\frac{A}{\pi}\left(1-\frac{C^X}{\pi^X}\right)E(Y|C^X=1, A, Z)-\frac{A}{\pi}\mu_1.
\end{split}
\end{align}
Assume that the covariates $\bm{Z}$, the treatment indicator $A$, the missingness indicator $C^X$ and the intermediate endpoint $X$ contain sufficient information to explain the missingness so that the missing at random (MAR) assumption, $Y\indep C^Y|A, \bm{Z}, C^X, X$ (Tsiatis, 2006), holds. All RAL estimators of $\mu_1$, under the assumption that $\pi^Y$ is known, have influence function
\begin{align}\label{phis3}
\left\{\frac{C^Y\varphi^{(1)}}{\pi^Y}+\left(1-\frac{C^Y}{\pi^Y}\right)L^{(2)}(A, \bm{Z}, C^X, X)\right\},
\end{align}
where $L^{(2)}(A, \bm{Z}, C^X, X)$ is an arbitrary function of $A$, $Z$, $C^X$ and $X$. 
From Theorem $10.2$ in Tsiatis (2006), for given $\varphi^{(1)}$, the optimal choice for $L^{(2)}(A, \bm{Z}, C^X, X)$ that leads to a semi-parametric efficient estimator under the above restrictions is given by $E\left(\varphi^{(1)}|C^Y=C^X=1, A, Z, X\right)$. Therefore, among the class of influence functions (\ref{phis3}), the one with the smallest variance is
\begin{align}\label{phi3}
\begin{split}
\varphi^{(2)} =&\frac{C^Y\varphi^{(1)}}{\pi^Y}+\left(1-\frac{C^Y}{\pi^Y}\right)E(\varphi^{(1)}|C^Y=C^X=1, A, Z, X)\\
=& \frac{C^YC^XA}{\pi^Y\pi^X\pi}Y+\frac{AC^X}{\pi\pi^X}\left(1-\frac{C^Y}{\pi^Y}\right)E(Y|C^Y=C^X=1, A, Z, X)\\
&+\frac{A}{\pi}\left(1-\frac{C^X}{\pi^X}\right)E(Y|C^X=1, A, Z)-\frac{A}{\pi}\mu_1.
\end{split}
\end{align}
This suggests estimating $\mu_1$ as the solution to the estimating equation
\begin{align*}
\sum_{i=1}^{n'}& \left\{ \frac{C^Y_iC^X_iA_i}{\hat{\pi}^Y\hat{\pi}^X\hat{\pi}}Y_i+\frac{A_iC^X_i}{\hat{\pi}\hat{\pi}^X}\left(1-\frac{C^Y_i}{\hat{\pi}^Y}\right)\hat{Y}_{1i} \right.\\
&\left.+\frac{A_i}{\hat{\pi}}\left(1-\frac{C^X_i}{\hat{\pi}^X}\right)\hat{Y}_{1i}'-\frac{A_i}{\hat{\pi}}\mu_1\right\}=0,
\end{align*}
where $n'$ is the number of recruited patients at the interim analysis, $\hat{Y}_{1i}=h_1(X_i, \bm{Z}_i, \bm{\hat{\eta}})$ with $E(Y|A=1, \bm{Z}, X)=h_1(X, \bm{Z}, \bm{\eta}_0)$, where $h_1(X, \bm{Z}, \bm{\eta})$ is a known function, evaluated at a parameter $\bm{\eta}$ with unknown population value $\bm{\eta}_0$,
and $\hat{Y}_{1i}'=f_1(\bm{Z}_i, \bm{\hat{\delta}})$ with $E(Y^*|A=1, \bm{Z})=f_1(\bm{Z},\bm{\delta}_0)$, where $f_1(\bm{Z}, \bm{\delta})$ is a known function, evaluated at a parameter $\bm{\delta}$ with unknown population value $\bm{\delta}_0$.
This yields
\begin{align*}
\hat{\mu}_{1} 
&=\frac{\sum_{i=1}^{n'} \left\{ \frac{C^Y_iC^X_iA_i}{\hat{\pi}^Y\hat{\pi}^X\hat{\pi}}Y_i+\frac{A_iC^X_i}{\hat{\pi}\hat{\pi}^X}\left(1-\frac{C^Y_i}{\hat{\pi}^Y}\right)\hat{Y}_{1i}+\frac{A_i}{\hat{\pi}}\left(1-\frac{C^X_i}{\pi^X}\right)\hat{Y}_{1i}'\right\}}{\sum_{i=1}^{n'} \frac{A_i}{\hat{\pi}}}\\
&=n'^{-1}\sum_{i=1}^{n'} \left\{ \frac{C^Y_iC^X_iA_i}{\hat{\pi}^Y\hat{\pi}^X\hat{\pi}}Y_i+\frac{A_iC^X_i}{\hat{\pi}\hat{\pi}^X}\left(1-\frac{C^Y_i}{\hat{\pi}^Y}\right)\hat{Y}_{1i}+\frac{A_i}{\hat{\pi}}\left(1-\frac{C^X_i}{\hat{\pi}^X}\right)\hat{Y}_{1i}'\right\},
\end{align*}
since $n'\hat{\pi}=\sum_{i=1}^{n'}A_i$. An alternative derivation is given in Appendix A.3.

\subsubsection*{Appendix A.2: Robustness of the Interim Estimator}
Under the considered recruiting strategy, we can assume independence of ($C^Y$, $C^X$, $C^Z$) and ($X$, $\bm{Z}$). Moreover, in absence of a time effect, random recruitment makes it quite plausible that $C^X_i \indep Y_i|\bm{Z}_i, A_i$ and $C^Y_i \indep Y_i|\bm{Z}_i, X_i, A_i, C^X_i=1$. Consequently, $\hat{\pi}^X$ and $\hat{\pi}^Y$ are consistent estimators for the conditional probabilities $P(C^X=1|A, \bm{Z})$ and $P(C^Y=1|C^X=1, A, X, \bm{Z})$ respectively. 
Additionally, $\hat{\pi}$ is a consistent estimator for $P(A=1)$. \\
Define the probability limits $\bm{\eta^*}=\text{plim}(\bm{\hat{\eta}})$ and $\bm{\delta^*}=\text{plim}(\bm{\hat{\delta}})$, which equal respectively the true values $\bm{\eta}_0$ and $\bm{\delta}_0$ when the working models $h_1(\bm{Z}, X; \bm{\eta})$  and $f_1(\bm{Z}; \bm{\delta})$ are correctly specified, but not necessarily otherwise. 
By the weak law of large numbers
and Slutsky's theorem, $\hat{\mu_1}$ estimates 
\begin{align*}
E&\left[  \frac{C^YC^XA}{\pi^Y\pi^X\pi}Y+\frac{AC^X}{\pi\pi^X}\left(1-\frac{C^Y}{\pi^Y}\right)h_1(X, \bm{Z}, \bm{\eta}^*)+\frac{A}{\pi}\left(1-\frac{C^X}{\pi^X}\right)f_1(\bm{Z}, \bm{\delta}^*)\right]\\
=& E\left[Y|A=1, C^Y=1, C^X=1 \right]+\\&E\left[ E\left\{\left(1-\frac{C^Y}{\pi^Y}\right)h_1(X, \bm{Z}, \bm{\eta}^*)  \middle|A=1, C^X=1, X, \bm{Z}\right\}\middle|A=1, C^X=1\right]+\\&
E\left[ E\left\{ \left(1-\frac{C^X}{\pi^X}\right)f_1(\bm{Z}, \bm{\delta}^*)\middle|A=1, \bm{Z}\right\}\middle|A=1\right]\\
=& E\left[Y|A=1, C^Y=1, C^X=1 \right]\\
=& E\left[Y|A=1\right]
\end{align*}
We therefore describe the proposed interim estimator as robust, in the sense that it is consistent even if the outcome models $h_1(X, \bm{Z}, \bm{\eta})$ and $f_1(\bm{Z}, \bm{\delta})$ for respectively $E(Y|A=1, X, \bm{Z})$ and $E(Y^*|A=1, \bm{Z})$ are misspecified (Tsiatis, $2008$).

\subsubsection*{Appendix A.3: Asymptotic Variance of the Interim Estimator}
Based on the derivation in Section 2.2, $\hat{\mu}_{1}$ can be written as
\begin{align}
\label{mu1}
\begin{split}
\hat{\mu}_{1} &= \frac{\sum_{i=1}^{n}A_i\left\{C^Y_iY_i+(1-C^Y_i)C^X_i\hat{Y}_{1i}+(1-C^X_i)\hat{Y}_{1i}'\right\}}{\sum_{i=1}^{n'} A_i}\\
&={n'}^{-1}\sum_{i=1}^{n'}\frac{A_i}{\hat{\pi}} \left\{C^Y_iY_i+(1-C^Y_i)C^X_i\hat{Y}_{1i}+(1-C^X_i)\hat{Y}_{1i}'\right\},
\end{split}
\end{align}
since $n'\hat{\pi}=\sum_{i=1}^{n'}A_i$. To obtain the variance of this estimator, one must take into account that predicted, and not observed, values $\hat{Y}_{1i}$ and $\hat{Y}_{1i}'$ are used in respectively cohort $2$ and $3$. By adding the (weighted) sum of the residual terms ${n'}^{-1}\sum_{i=1}^{n'}(A_i/\hat{\pi})\left((1-\hat{\pi}^X\hat{\pi}^Y)/(\hat{\pi}^X\hat{\hat{\pi}}^Y)\right)C^X_iC^Y_i(Y_i-\hat{Y}_{1i})$ and ${n'}^{-1}\sum_{i=1}^{n'}(A_i/\hat{\pi})\left((1-\hat{\pi}^X)/\hat{\pi}^X\right)C^X_i(\hat{Y}_{1i}-\hat{Y}_{1i}')$ in Equation (\ref{mu1}), we correct for the estimation of the parameters in the canonical generalized linear models used to estimate respectively $\hat{Y}_{1i}$ and $\hat{Y}_{1i}'$ (Tsiatis, $2006$). 
Since we are dealing with canonical generalized linear models, the sum of both residuals sum to zero and therefore adding these sums to the equations does not change the estimators $\hat{\mu}_1$. Rearranging the terms yields
\begin{align*}
\hat{\mu}_{1} &={n'}^{-1}\sum_{i=1}^{n'}\frac{A_i}{\hat{\pi}}\left(\frac{C^Y_iC^X_i}{\hat{\pi}^Y\hat{\pi}^X}(Y_i-\hat{Y}_{1i})+\frac{C^X_i}{\hat{\pi}^X}(\hat{Y}_{1i}-\hat{Y}_{1i}')+\hat{Y}_{1i}'\right).
\end{align*}
Since the asymptotic behavior of an estimator is fully determined by
its influence function, it suffices to focus on
the influence function when discussing the estimator's variance. Define
\begin{align*}
\phi_1(X, \bm{Z}; P_1, \bm{\eta}, \bm{\delta})=&\frac{A}{\pi}\left[\frac{C^YC^X}{\pi^Y\pi^X}\left\{Y-h_1(X, \bm{Z},\bm{\eta})\right\}
+\frac{C^X}{\pi^X}\left\{h_1(X, \bm{Z},\bm{\eta})-f_1(\bm{Z},\bm{\delta})\right\}\right.\\&\left.+f_1(\bm{Z},\bm{\delta})-P_1\vphantom{\frac{1-A_i}{1-\pi}}\right],
\end{align*}
where $E(Y|A=1, X, \bm{Z}) = h_1(\bm{Z}, X; \bm{\eta}_0)$ and $E(Y^*|A=1, \bm{Z}) = f_1(\bm{Z}; \bm{\delta}_0)$. When the working models $h_1(\bm{Z}, X; \bm{\eta})$  and $f_1(\bm{Z}; \bm{\delta})$ are correctly specified, then the probability limits $\bm{\eta}^*=\text{plim}(\hat{\bm{\eta}})$ and $\bm{\delta}^*=\text{plim}(\hat{\bm{\delta}})$ equal respectively $\bm{\eta}_0$ and $\bm{\delta}_0$, but not necessarily otherwise. 
Under the assumption that $\hat{\bm{\eta}}$, $\hat{\bm{\delta}}$, $\hat{\pi}$, $\hat{\pi}^X$ and $\hat{\pi}^Y$ are the solution to respectively the estimating equations ${n'}^{-1}\sum_{i=1}^{n'}\bm{U}_\eta(X_i, \bm{Z}_i; \bm{\eta})=\bm{0}$, ${n'}^{-1}\sum_{i=1}^{n'}\bm{U}_\delta(\bm{Z_i}; \bm{\delta})=\bm{0}$, ${n'}^{-1}\sum_{i=1}^{n'}\bm{U}_\pi(\pi)=0$, ${n'}^{-1}\sum_{i=1}^{n'}\bm{U}_{\pi^X}(\pi^X)=0$ and ${n'}^{-1}\sum_{i=1}^{n'}\bm{U}_{\pi^Y}(\pi^Y)=0$, the influence function (Tsiatis, $2006$) is given by
\begin{align*}
\tilde{\phi}_1(X, \bm{Z}; P_1, \bm{\eta}^*, \bm{\delta}^*) =& \phi_1(X, \bm{Z}; P_1, \bm{\eta}^*, \bm{\delta}^*)\\
&-E\left\{\frac{\partial \phi_1}{\partial \bm{\eta}^T}(X, \bm{Z}; P_1, \bm{\eta}^*, \bm{\delta}^*) \right\}E^{-1}\left\{\frac{\partial \bm{U}_\eta}{\partial \bm{\eta}^T}(X, \bm{Z}; \bm{\eta}^*)\right\}\bm{U}_\eta(X, \bm{Z}; \bm{\eta}^*)\\
&-E\left\{\frac{\partial \phi_1}{\partial \bm{\delta}^T}(X, \bm{Z}; P_1, \bm{\eta}^*, \bm{\delta}^*) \right\}E^{-1}\left\{\frac{\partial \bm{U}_\delta}{\partial \bm{\delta}^T}(\bm{Z}; \bm{\delta}^*)\right\}\bm{U}_\delta(\bm{Z}; \bm{\delta}^*)\\
&-E\left\{\frac{\partial \phi_1}{\partial \pi}(X, \bm{Z}; P_1, \bm{\eta}^*, \bm{\delta}^*) \right\}E^{-1}\left\{\frac{\partial \bm{U}_\pi}{\partial \pi}\right\}\bm{U}_\pi\\
&-E\left\{\frac{\partial \phi_1}{\partial \pi^X}(X, \bm{Z}; P_1, \bm{\eta}^*, \bm{\delta}^*) \right\}E^{-1}\left\{\frac{\partial \bm{U}_\pi}{\partial {\pi^X}}\right\}\bm{U}_{\pi}^X\\
&-E\left\{\frac{\partial \phi_1}{\partial \pi^Y}(X, \bm{Z}; P_1, \bm{\eta}^*, \bm{\delta}^*) \right\}E^{-1}\left\{\frac{\partial \bm{U_\pi}}{\partial {\pi^Y}}\right\}\bm{U}_{\pi}^Y.
\end{align*}
These, 
\begin{align*}
E\left\{\frac{\partial \phi_1}{\partial \bm{\eta}^T}(X, \bm{Z}; P_1, \bm{\eta}^*, \bm{\delta}^*)\right\}=&E\left\{\frac{\partial h_1}{\partial \bm{\eta}^T}(X, \bm{Z},\bm{\eta}^*)\frac{AC^X}{\pi\pi^X}\left(1- \frac{C^Y}{\pi^Y}\right)\right\}=\bm{0},
\end{align*}
and
\begin{align*}
E\left\{\frac{\partial \phi_1}{\partial \bm{\delta}^T}(X, \bm{Z}; P_1, \bm{\eta}^*, \bm{\delta}^*)\right\}=&E\left\{\frac{\partial f_1}{\partial \bm{\delta}^T}(\bm{Z},\bm{\delta}^*)\frac{A}{\pi}\left(1- \frac{C^X}{\pi^X}\right)\right\}=\bm{0},
\end{align*}
since the propensity scores $\pi^Y$ and $\pi^X$ respectively are guaranteed to be correctly specified under random recruitment. Next, 
\begin{align*}
E\left\{\frac{\partial \phi_1}{\partial \pi}(X, \bm{Z}; P_1, \bm{\eta}^*, \bm{\delta}^*) \right\}=&E\left\{\frac{-A}{\pi^2}\left(\frac{C^YC^X}{\pi^Y\pi^X}\left(Y-h_1(X, \bm{Z},\bm{\eta})\right)
+\frac{C^X}{\pi^X}\left(h_1(X, \bm{Z},\bm{\eta})-f_1(\bm{Z},\bm{\delta})\right)\right.\right.\\&\left.\left.+f_1(\bm{Z},\bm{\delta})-P_1\vphantom{\frac{1-A_i}{1-\pi}}\right)\right\}\\
=&E\left\{\frac{-A}{\pi^2}\left(h_1(X, \bm{Z},\bm{\eta})\frac{C^X}{\pi^X}\left(1-\frac{C^Y}{\pi^Y}\right)
+f_1(\bm{Z},\bm{\delta})\left(1-\frac{C^X}{\pi^X}\right)
\right.\right.\\&\left.\left.+\frac{C^YC^X}{\pi^Y\pi^X}Y-P_1\vphantom{\frac{1-A_i}{1-\pi}}\right)\right\}=0,
\end{align*}
by the fact that the propensity scores $\pi^Y$ and $\pi^X$ are guaranteed to be correctly specified under random recruitment and the fact that $\sum_{i=1}^{n'}(A_iC^X_iC^Y_i/(\pi\pi^X\pi^Y))Y_i$ is an unbiased estimator for $P_1$. Finally,
\begin{align*}
E\left\{\frac{\partial \phi_1}{\partial \pi^X}(X, \bm{Z}; P_1, \bm{\eta}^*, \bm{\delta}^*) \right\}=
&E\left\{\frac{-1}{\pi^X}\left(\frac{AC^XC^Y}{\pi\pi^X\pi^Y}\left(Y-h_1(X, \bm{Z},\bm{\eta})\right)
+\frac{AC^X}{\pi\pi^X}\left(h_1(X, \bm{Z},\bm{\eta})-f_1(\bm{Z},\bm{\delta})\right)\right)\right\}\\=&0,
\end{align*}
and
\begin{align*}
E\left\{\frac{\partial \phi_1}{\partial \pi^Y}(X, \bm{Z}; P_1, \bm{\eta}^*, \bm{\delta}^*) \right\}=&E\left\{\frac{-1}{\pi^Y}\frac{AC^XC^Y}{\pi\pi^X\pi^Y}\left(Y-h_1(X, \bm{Z},\bm{\eta})\right)\right\}=0,
\end{align*}
because of the way $\bm{\eta}$ and $\bm{\delta}$ are estimated. Therefore, the influence function reduces to
$
\phi_1(X, \bm{Z}; P_1, \bm{\eta}^*, \bm{\delta}^*)$.
Similar reasonings lead to the estimator 
\begin{align*}
\hat{\mu}_{0} &={n'}^{-1}\sum_{i=1}^{n'}\frac{1-A_i}{1-\hat{\pi}}\left(\frac{C^Y_iC^X_i}{\hat{\pi}^Y\hat{\pi}^X}(Y_i-\hat{Y}_{0i})+\frac{C^X_i}{\hat{\pi}^X}(\hat{Y}_{0i}-\hat{Y}_{0i}')+\hat{Y}_{0i}'\right),
\end{align*}
with influence function
\begin{align*}
\phi_0(X, \bm{Z}; P_0, \bm{\zeta}, \bm{\nu}) =&\frac{1-A}{1-\pi}\left[\frac{C^YC^X}{\pi^Y\pi^X}\left\{Y-h_0(X, \bm{Z},\bm{\zeta})\right\}+\frac{C^X}{\pi^X}\left\{h_0(X, \bm{Z},\bm{\zeta})-f_0(\bm{Z},\bm{\nu})\right\}\right.\\
&\left.+f_0(\bm{Z},\bm{\nu})-P_0\vphantom{\frac{1-A_i}{1-\pi}}\right],
\end{align*}
where $E(Y|A=0, X, \bm{Z}) = h_0(\bm{Z}, X; \bm{\zeta}_0)$ and $E(Y^*|A=1, \bm{Z}) = f_0(\bm{Z}; \bm{\nu}_0)$.
The asymptotic variance of the estimator $\hat{\mu}_1-\hat{\mu}_0$, denoted by $s^2$, can then be straightforwardly estimated as one over $n'$ times the sample variance of the values 
\begin{align*}
\phi_1(X, \bm{Z}_i; \hat{\mu}_{1}, \hat{\bm{\eta}}, \hat{\bm{\delta}})-\phi_0(X, \bm{Z}_i; \hat{\mu}_{0}, \hat{\bm{\zeta}}, \hat{\bm{\nu}}).
\end{align*}
Note that further efficiency can sometimes be gained by estimating the probabilities $\pi$, $\pi^X$ and $\pi^Y$ conditional on the covariates. We can then use a weighted regression, based on these weights, to impute the primary endpoint values in cohort $2$ and $3$.
\subsection*{Appendix B: Independent Increments Property}
Define $S_t$ as
\begin{align*}
&\frac{A}{\pi}\left\{\frac{C^YC^X}{\pi^Y\pi^X}Y+\frac{C^X}{\pi^X}\left(1-\frac{C^Y}{\pi^Y}\right)h_1(X, \bm{Z}, \bm{\eta}_0)+\left(1-\frac{C^X}{\pi^X}\right)f_1(\bm{Z}, \bm{\delta}_0)-\mu_1\right\}\\
&-\frac{1-A}{1-\pi}\left\{\frac{C^YC^X}{\pi^Y\pi^X}Y+\frac{C^X}{\pi^X}\left(1-\frac{C^Y_i}{\pi^Y}\right)h_0(X, \bm{Z}, \bm{\zeta}_0)+\left(1-\frac{C^X}{\pi^X}\right)f_0(\bm{Z}, \bm{\nu}_0)-\mu_0\right\},
\end{align*}
and $S_{t,i}$ as
\begin{align*}
&\frac{A_i}{\hat{\pi}}\left\{\frac{C^Y_iC^X_i}{\hat{\pi}^Y_i\hat{\pi}^X}Y+\frac{C^X_i}{\hat{\pi}^X}\left(1-\frac{C^Y_i}{\hat{\pi}^Y}\right)h_1(X_i, \bm{Z}_i, \hat{\bm{\eta}})+\left(1-\frac{C^X_i}{\hat{\pi}^X}\right)f_1(\bm{Z}_i, \hat{\bm{\delta}})-\mu_1\right\}\\
&-\frac{1-A_i}{1-\hat{\pi}}\left\{\frac{C^Y_iC^X_i}{\hat{\pi}^Y\hat{\pi}^X}Y+\frac{C^X_i}{\hat{\pi}^X}\left(1-\frac{C^Y_i}{\hat{\pi}^Y}\right)h_0(X_i, \bm{Z}_i, \hat{\bm{\zeta}})+\left(1-\frac{C^X_i}{\hat{\pi}^X}\right)f_0(\bm{Z}_i, \hat{\bm{\nu}})-\mu_0\right\},
\end{align*}
where $C^Y_i$ and $C^X_i$ are evaluated at interim time $t$.
Then
\begin{align*}
\sqrt{n'}\left\{(\hat{\mu}_1-\hat{\mu}_0)-(\mu_1-\mu_0)\right\}={n'}^{-1/2}\sum_{i=1}^{n'}S_{t, i}.
\end{align*}
Therefore, $\hat{\mu}_1-\hat{\mu}_0$ is an asymptotically linear estimator of $\mu_1-\mu_0$ whose influence function is given by $\phi_i=S_{t, i}$ for the $i$-th subject. Suppose that we are interested in testing $H_0: \mu_1-\mu_0=\mu^0_1-\mu^0_0$, then $\mu_j$ in the previous expression is replaced by $\mu_j^0$ ($j\in\{0,1\}$):
\begin{align*}
{n'}^{-1/2}\sum_{i=1}^{n'}S_{t, i}&=\sqrt{n'}\left\{(\hat{\mu}_1-\hat{\mu}_0)-(\mu_1-\mu_0)\right\}\\
&=\sqrt{n'}SE(\hat{\mu}_1-\hat{\mu}_0)\left\{(\hat{\mu}_1-\hat{\mu}_0)-(\mu_1-\mu_0)\right\}/SE(\hat{\mu}_1-\hat{\mu}_0)\\
&=\sqrt{n'}SE(\hat{\mu}_1-\hat{\mu}_0)Z_t\\
&=\hat{\mathcal{I}}_t^{-1/2}Z_t,
\end{align*}
with $SE(\hat{\mu}_1-\hat{\mu}_0)$ the estimated standard error and $\hat{\mathcal{I}}_t$ the observed information which equals $1/(SE(\hat{\mu}_1-\hat{\mu}_0)^2n')$. 
Since $\mathcal{I}_t/\hat{\mathcal{I}}_t=SE(\hat{\mu}_1-\hat{\mu}_0)^2n'/\sigma^2$, with $\sigma^2=\text{Var}(S_t)$ and $\mathcal{I}_t$ the expected information, converges in probability to $1$, we obtain that 
$
Z_t = {n'}^{-1/2}\sum_{i=1}^{n'}\mathcal{I}_t^{1/2}S_{t,i}+o_p(1)
$.
Moreover, note that $\hat{\mathcal{I}}_t^{1/2}Z_t$ is equal to ${n'}^{-1/2}\sum_{i=1}^{n'}\mathcal{I}_tS_{t,i}+o_p(1)$. 

Since the test statistic at the end of the study, \begin{align*}
Z_1=\frac{\hat{P}_1-\hat{P}_0}{\sqrt{\frac{\hat{P}_1(1-\hat{P}_1)}{n_1}+\frac{\hat{P}_0(1-\hat{P}_0)}{n_0}}}
\end{align*}
where $\hat{P}_1=\sum_{i=1}^{n}A_iY_i/\sum_{i=1}^{n}A_i$ and $\hat{P}_0=\sum_{i=1}^{n}(1-A_i)Y_i/\sum_{i=1}^{n}(1-A_i)$, is a special case of the interim test statistic (with $C^X=C^Y=1$ for all patients and $n'=n$), similar results hold for $Z_1$. Now consider a second interim analysis at time $s$ and define $S_s$ in a similar way as $S_t$ but with $C^X$ and $C^Y$ evaluated at time $s$.
To prove the independent increments assumption, it is sufficient to establish that $E[S_t\mathcal{I}_t(S_s\mathcal{I}_s-S_t\mathcal{I}_t)]=0$, with $t$ and $s$ two arbitrary timepoints where $0<t<s$.
To do this, note that
\begin{align*}
E\left[S_1(S_1-S_t)\right]=& E\left[\frac{A}{\pi^2}(Y-\mu_1)\left\{(Y-\mu_1)-\frac{C^YC^X}{\pi^Y\pi^X}Y-\frac{C^X}{\pi^X}\left(1-\frac{C^Y}{\pi^Y}\right)h_1(X, \bm{Z}, \bm{\eta}^*)\right. \right.\\
&-\left.\left.\left(1-\frac{C^X}{\pi^X}\right)f_1(\bm{Z}, \bm{\delta}^*)+\mu_1\right\} \right]\\
&- E\left[\frac{1-A}{(1-\pi)^2}(Y-\mu_0)\left\{(Y-\mu_0)-\frac{C^YC^X}{\pi^Y\pi^X}Y-\frac{C^X}{\pi^X}\left(1-\frac{C^Y}{\pi^Y}\right)h_0(X, \bm{Z}, \bm{\zeta}^*)\right. \right.\\
&-\left.\left.\left(1-\frac{C^X}{\pi^X}\right)f_0(\bm{Z}, \bm{\nu}^*)+\mu_0\right\}\right]\\
=&\pi^{-1} E\left[Y\left\{Y-\frac{C^YC^X}{\pi^Y\pi^X}Y-\frac{C^X}{\pi^X}\left(1-\frac{C^Y}{\pi^Y}\right)h_1(X, \bm{Z}, \bm{\eta}^*)\right.\right.\\
&-\left.\left.\left(1-\frac{C^X}{\pi^X}\right)f_1(\bm{Z}, \bm{\delta}^*)\right\} \middle|A=1\right]\\
&-(1-\pi)^{-1} E\left[Y\left\{Y-\frac{C^YC^X}{\pi^Y\pi^X}Y-\frac{C^X}{\pi^X}\left(1-\frac{C^Y}{\pi^Y}\right)h_0(X, \bm{Z}, \bm{\zeta}^*)\right.\right.\\
&-\left.\left.\left(1-\frac{C^X}{\pi^X}\right)f_0(\bm{Z}, \bm{\nu}^*)\right\}\middle|A=0\right]\\
=&\pi^{-1} E\left[Y^2-\frac{C^YC^X}{\pi^Y\pi^X}Y^2-\frac{C^X}{\pi^X}\left(1-\frac{C^Y}{\pi^Y}\right)h_1(X, \bm{Z}, \bm{\eta}^*)Y\right.\\
&-\left.\left(1-\frac{C^X}{\pi^X}\right)f_1(\bm{Z}, \bm{\delta}^*)Y \middle|A=1\right]\\
&-(1-\pi)^{-1} E\left[Y^2-\frac{C^YC^X}{\pi^Y\pi^X}Y^2-\frac{C^X}{\pi^X}\left(1-\frac{C^Y_i}{\pi^Y}\right)h_0(X, \bm{Z}, \bm{\zeta}^*)Y\right.\\
&-\left.\left(1-\frac{C^X}{\pi^X}\right)f_0(\bm{Z}, \bm{\nu}^*)Y\middle|A=0\right]\\
=&\pi^{-1} E\left[Y^2-\frac{C^YC^X}{\pi^Y\pi^X}Y^2 \middle|A=1\right]-(1-\pi)^{-1} E\left[Y^2-\frac{C^YC^X}{\pi^Y\pi^X}Y^2\middle|A=0\right]\\
=& 0.
\end{align*}
In a similar way, it can be shown that $E\left[S_s(S_s-S_t)\right]=0$.
Consequently, $E\left[S_sS_t\right]=E\left[S_s^2\right]$. With this in mind,
\begin{align*}
E[S_t\mathcal{I}_t(S_s\mathcal{I}_s-S_t\mathcal{I}_t)]&=E[S_t/E[S_t^2](S_s/E[S_s^2]-S_t/E[S_t^2])]\\
&=E[S_tS_s]/(E[S_t^2]E[S_s^2])-E[S_t^2]/E[S_t^2]^2\\
&=E[S_s^2]/(E[S_t^2]E[S_s^2])-1/E[S_t^2]\\
&=0,
\end{align*}
since $\mathcal{I}_t=1/E\left[S_t^2\right]$ and $\mathcal{I}_s=1/E\left[S_s^2\right]$.
Because the basic Brownian motion joint distribution structure is satisfied, the proposed test statistics can be imbedded in the conditional power method.

For the sake of completeness, we show in a similar way that the $B$-value $B_t=Z_t\sqrt{t}$ at information time $t$  is independent of the increment $B_1-B_t=Z_1-Z_t\sqrt{t}$. To establish this, it is sufficient to prove that $Cov[S_t\mathcal{I}_t^{1/2}\sqrt{\mathcal{I}_t/\mathcal{I}_1},S_1\mathcal{I}_1^{1/2}-S_t\mathcal{I}_t^{1/2}\sqrt{\mathcal{I}_t/\mathcal{I}_1}]=0$,  since $t=\frac{\mathcal{I}_t}{\mathcal{I}_1}$.
\begin{align*}
Cov[S_t\mathcal{I}_t^{1/2}\sqrt{\mathcal{I}_t/\mathcal{I}_1},S_1\mathcal{I}_1^{1/2}-S_t\mathcal{I}_t^{1/2}\sqrt{\mathcal{I}_t/\mathcal{I}_1}]&=
E\left[\frac{S_t\mathcal{I}_t}{\sqrt{\mathcal{I}_1}}\left(S_1\mathcal{I}_1^{1/2}-\frac{S_t\mathcal{I}_t}{\sqrt{\mathcal{I}_1}}\right)\right]\\
&=E\left[S_tS_1\right]\mathcal{I}_t-E\left[S_t^2\right]\mathcal{I}_t^2/\mathcal{I}_1\\
&=E\left[S_1^2\right]\mathcal{I}_t-E\left[S_t^2\right]\mathcal{I}_t^2/\mathcal{I}_1,
\end{align*}
since $\mathcal{I}_t=1/E\left[S_t^2\right]$ and $\mathcal{I}_1=1/E\left[S_1^2\right]$, this yields
\begin{align*}
Cov[S_t\mathcal{I}_t^{1/2}\sqrt{\mathcal{I}_t/\mathcal{I}_1},S_1\mathcal{I}_1^{1/2}-S_t\mathcal{I}_t^{1/2}\sqrt{\mathcal{I}_t/\mathcal{I}_1}]&=E\left[S_1^2\right]\mathcal{I}_t-E\left[S_t^2\right]\mathcal{I}_t^2/\mathcal{I}_1\\
&=E\left[S_1^2\right]/E\left[S_t^2\right]-E\left[S_t^2\right]E\left[S_1^2\right]/E\left[S_t^2\right]^2\\
&=E\left[S_1^2\right]/E\left[S_t^2\right]-E\left[S_1^2\right]/E\left[S_t^2\right]\\
&=0.
\end{align*}
Because the basic Brownian motion joint distribution structure is satisfied, the proposed test statistics can be imbedded in the conditional power method.\\[2ex]

\subsection*{Appendix C: Sample Size Reassessment}
This section focuses on the sample size reassessment based on conditional power arguments.
\subsubsection*{Appendix C.1: Independent Test Statistics}
Since $Z_t\sqrt{t}\indep (Z_1-Z_t\sqrt{t})$, it is intuitively clear that the second stage test statistic should be based on the independent increment $Z_1-Z_t\sqrt{t}$. 
Moreover, this independence leads to the fact that $Var(Z_1-Z_t\sqrt{t})=1-t$ since $Var(Z_1)=Var(Z_1-Z_t\sqrt{t})+Var(Z_t\sqrt{t})$.
We therefore suggest the normalized statistic $(Z_1-\sqrt{t}Z_t)/\sqrt{1-t}$. It then follows directly from Appendix B that this second stage test statistic is asymptotically independent of the first stage test statistic $Z_t$, since:
\begin{align*}
Cov[Z_t,(Z_1-\sqrt{t}Z_t)/\sqrt{1-t}]&=Cov[\sqrt{t}Z_t,Z_1-\sqrt{t}Z_t]/(\sqrt{1-t}\sqrt{t})
=0.
\end{align*}

\subsubsection*{Appendix C.2: Theoretical Derivation of the Sample Size Reassessment Formula}
Define $\theta=\Delta\sqrt{n}/\sigma$, with $\Delta$ the treatment effect $P_1-P_0$ used for powering the study and $\sigma^2$ the common variance in both treatment arms.
Introduce the artifical \textit{sample size} at the interim analysis as 
$n_B=t\cdot n$, let $n_2$ correspond to the second-stage sample size and denote the adjusted total sample size as $\tilde{n}$. Writing the assumption on the second stage data in terms of the second stage sample size, the conditional power equations under the design effect can be rewritten as 
\begin{align*}
CP_t(\theta) &=  1-\Phi\left(\frac{z_{1-\alpha}-Z^{(1)}\sqrt{w}}{\sqrt{1-w}}-\theta\sqrt{\frac{n_2}{n}}\right)\\
&= 1-\Phi\left(\frac{z_{1-\alpha}-Z^{(1)}\sqrt{w}}{\sqrt{1-w}}-\frac{\theta}{\sqrt{n}}\sqrt{n_2}\right),
\end{align*}
where the weights $w$ and $1-w$ ($0\leq w\leq 1$) are the pre-specified weights corresponding with the first and second stage test statistic of the combination test.
Since $\theta/\sqrt{n}=\Delta/\sigma$, $n_2$ is the only part of the equation that is not fixed at the time of the interim analysis. The target conditional power, $1-\beta$, can therefore be achieved by the choice of the second stage \textit{sample size} $n_2=\tilde{n}-n_B$; choosing $\tilde{n}$ such that the conditional power equals $1-\beta$, gives
\begin{align*}
& &1-\beta &= 1-\Phi\left(\frac{z_{1-\alpha}-Z^{(1)}\sqrt{w}}{\sqrt{1-w}}-\frac{\theta}{\sqrt{n}}\sqrt{\tilde{n}-n_B}\right)\\
&\Longleftrightarrow & \Phi^{-1}(\beta)  &= \frac{z_{1-\alpha}-Z^{(1)}\sqrt{w}}{\sqrt{1-w}}-\frac{\theta}{\sqrt{n}}\sqrt{\tilde{n}-n_B}\\
&\Longleftrightarrow & \sqrt{\tilde{n}-n_B} &= \frac{\frac{z_{1-\alpha}-Z^{(1)}\sqrt{w}}{\sqrt{1-w}}-\Phi^{-1}(\beta)}{\theta/\sqrt{n}}\\
&\Longleftrightarrow & \tilde{n}-n_B &= \max\left\{0, \left(\frac{\frac{z_{1-\alpha}-Z^{(1)}\sqrt{w}}{\sqrt{1-w}}-\Phi^{-1}(\beta)}{\theta/\sqrt{n}}\right)\right\}^2.
\end{align*}
If $\tilde{n}\leq n'$, with $n'$ the number of recruited patients at the interim analysis, then no extra subjects are recruited. Note that a second test statistic is still being calculated after the primary endpoint is available for all $n'$ recruited patients.
In practice, the sample size reassessment is often done assuming the observed effect for the primary endpoint data in the remainder of the study. Similar reasonings lead to
\begin{align*}
& &1-\beta &= 1-\Phi\left(\frac{z_{1-\alpha}-Z^{(1)}\sqrt{w}}{\sqrt{1-w}}-\frac{Z^{(1)}/\sqrt{t}}{\sqrt{n}}\sqrt{\tilde{n}-n_B}\right)\\
&\Longleftrightarrow & \Phi^{-1}(\beta)  &= \frac{z_{1-\alpha}-Z^{(1)}\sqrt{w}}{\sqrt{1-w}}-\frac{Z^{(1)}/\sqrt{t}}{\sqrt{n}}\sqrt{\tilde{n}-n_B}\\
&\Longleftrightarrow & \sqrt{\tilde{n}-n_B} &= \frac{\frac{z_{1-\alpha}-Z^{(1)}\sqrt{w}}{\sqrt{1-w}}-\Phi^{-1}(\beta)}{(Z^{(1)}/\sqrt{t})/\sqrt{n}}\\
&\Longleftrightarrow & \tilde{n}-n_B &= \max\left\{0, \left(\frac{\frac{z_{1-\alpha}-Z^{(1)}\sqrt{w}}{\sqrt{1-w}}-\Phi^{-1}(\beta)}{(Z^{(1)}/\sqrt{t})/\sqrt{n}}\right)\right\}^2.
\end{align*}

\subsection*{Appendix D: Theoretical Derivation of the Blinded Information Fraction}
The blinded estimate of the variance of the estimator at the end of the study equals 
\begin{align*}
Var\left[\frac{A}{\pi}Y-\frac{1-A}{1-\pi}Y-\left(\frac{A}{\pi}\mu-\frac{1-A}{1-\pi}\mu\right)\right]&=
Var\left[\left(\frac{A}{\pi}-\frac{1-A}{1-\pi}\right)(Y-\mu)\right]\\
&=
Var\left[\frac{A-\pi}{(1-\pi)\pi}(Y-\mu)\right]\\
&=
((1-\pi)\pi)^{-1}E\left[(Y-\mu)^2\right],
\end{align*} 
with $\mu$ the probability of success over the two treatment arms.
However, during monitoring of the trial, the primary endpoint is not observed for all patients. Nevertheless, we can calculate $(Y_i-\hat{\mu})^2$ for each patient $i$ for whom $Y_i$ is observed ($C^Y_i=1$) with $\hat{\mu}=\sum_{i=1}^{n}C^Y_iY_i/\sum_{i=1}^{n}C^Y_i$. Under the assumption that $Y_i\indep C^Y_i$, the variance is estimated as $1/(n(1-\pi)\pi)$ times the sample mean of $(Y-\hat{\mu})^2$ over the patients with observed $Y$ data.

The variance of the blinded estimator obtained during monitoring of the trial on the other hand equals
\begin{align*}
Var&\left[
\frac{AC^XC^Y}{\pi\pi^X\pi^Y}\left(Y-E(Y|C^Y=1, Z, X)\right)+\frac{AC^X}{\pi\pi^X}\left(E(Y|C^Y=1, Z, X)-E(Y|C^X=1, Z)\right)\right.\\
&+\frac{A}{\pi}E(Y|C^X=1, Z)
-\frac{(1-A)C^XC^Y}{(1-\pi)\pi^X\pi^Y}\left(Y-E(Y|C^Y=1, Z, X)\right)\\
&-\frac{(1-A)C^X}{(1-\pi)\pi^X}\left(E(Y|C^Y=1, Z, X)-E(Y|C^X=1, Z)\right)
-\frac{1-A}{1-\pi}E(Y|C^X=1, Z)\\
&\left.-\frac{A}{\pi}\mu+\frac{1-A}{1-\pi}\mu\right]\\
=&Var\left[
\frac{C^XC^Y}{\pi^X\pi^Y}\left\{A/\pi-(1-A)/(1-\pi)\right\}\left(Y-E(Y|C^Y=1, Z, X)\right)\right.\\
&+\frac{C^X}{\pi^X}\left\{A/\pi-(1-A)/(1-\pi)\right\}\left(E(Y|C^Y=1, Z, X)-E(Y|C^X=1, Z)\right)\\
&\left.+\left\{A/\pi-(1-A)/(1-\pi)\right\}E(Y|C^X=1, Z)-\left\{A/\pi-(1-A)/(1-\pi)\right\}\mu\vphantom{\frac{A}{\pi}}\right]\\
=&((1-\pi)\pi)^{-1}E\left[\left(
\frac{C^XC^Y}{\pi^X\pi^Y}\left(Y-E(Y|C^Y=1, Z, X)\right)\right.\right.\\
&+\left.\left.\frac{C^X}{\pi^X}\left(E(Y|C^Y=1, Z, X)-E(Y|C^X=1, Z)\right)
+E(Y|C^X=1, Z)-\mu\right)^2\right].
\end{align*} 
Thus, the variance is estimated as $1/(n'(1-\pi)\pi)$ times the sample mean of the values 
\begin{align*}
&\left(
\frac{C^X_iC^Y_i}{\pi^X\pi^Y}\left(Y_i-E(Y_i|C^Y_i=1, Z_i, X_i)\right)
+\frac{C^X_i}{\pi^X}\left(E(Y_i|C^Y_i=1, Z_i, X_i)-E(Y_i|C^X_i=1, Z_i)\right)\right.\\
&\left.+E(Y_i|C^X_i=1, Z_i)-\hat{\mu}\vphantom{\frac{1-A_i}{1-\pi}}\right)^2
\end{align*} 
over all $n'$ recruited patients ($C^Z_i=1$).\\

\subsection*{Appendix E: $\bm{R}$-squared}\label{rsquared}

We define the total R-squared, which quantifies the proportion of variation in $Y$ explained by $X$ and $\bm{Z}$, as
\begin{align*}
R^2=\frac{\hat{Var}[\hat{\beta}_0+\hat{\beta}_1X+\hat{\bm{\beta}}_2\bm{Z}]}{\hat{Var}[\hat{\beta}_0+\hat{\beta}_1X+\hat{\bm{\beta}}_2\bm{Z}]+\pi^2/4},
\end{align*}
with $\hat{\beta}_0$, $\hat{\beta}_1$ and $\hat{\bm{\beta}}_2$ the estimated parameters in a logistic regression model for $Y$ given $X$ and $\bm{Z}$, $E(Y|X, \bm{Z})=\text{logit}^{-1}(\beta_0+\beta_1X+\bm{\beta}_2\bm{Z})$ and $\hat{Var}(\cdot)$ the empirical variance. 
The proportion of variation explained by $X$ cannot be directly derived from one logistic regression since the value of $X$ depends on $\bm{Z}$. We therefore denote the estimated prediction $\hat{P}(X_i=1|\bm{Z}_i)$ from a logistic regression model with regressors $\bm{Z}_i$ as $\hat{Q}(\bm{Z}_i)$. We then define the proportion of variation in $Y$ explained by $X$ as
\begin{align*}
R^2_X=\frac{\hat{\beta}_1^2Var[X-\hat{Q}(\bm{Z})]}{\hat{Var}[\hat{\beta}_0+\hat{\beta}_1X+\hat{\bm{\beta}}_2\bm{Z}]+\pi^2/4}.
\end{align*}
Note that we cancel out the variation in $Y$ explained by $X$ via $\bm{Z}$ by incorporating $Var[X-\hat{Q}(\bm{Z})]$ to obtain the variation uniquely explained by $X$. We define the proportion of variation explained by $\bm{Z}$ is given by 
\begin{align*}
R^2_Z=\frac{Var[\hat{\beta}_1\hat{Q}(\bm{Z})+\hat{\bm{\beta}}_2\bm{Z}]}{\hat{Var}[\hat{\beta}_0+\hat{\beta}_1X+\hat{\bm{\beta}}_2\bm{Z}]+\pi^2/4},
\end{align*}
which guarantees that $R^2=R^2_X+R^2_Z$.

\subsection*{Appendix F: Tables}
\begin{table}
	\begin{threeparttable}
		\caption{\label{ssr_monit_obs}Operating characteristics of a trial with sample size re-assessment based on conditional power as futility stopping rule at the time $50\%$ of the total information is obtained. Results are based on $10,000$ ($10,0000$ for null hypothis) Monte Carlo simulations.}
		\centering
		\begin{tabular}{l l c c c c }
			\hline	\hline
			&&\multicolumn{4}{c}{Treatment Effect}\\	
			&&$0$& $0.09$  & $0.13$ & $0.16$\\ 
			\hline
			&Power One Stage Trial&$2.62\%$&$58.67\%$  &$90.52\%$&$98.73\%$\\ \hline
			\textbf{Proposal}&Average $\#$ Days&$970$&$951$ &$942$&$934$\\
			&FS&$59.9\%$&$9.5\%$ &$1.9\%$&$0.3\%$\\
			&Power Loss&$0.08\%$& $0.78\%$&$0.56\%$&$0.16\%$\\
			&Power No SSR&$ 2.54\%$& $57.89\%$ &$89.96\%$&$98.57\%$\\
			&$Q_0$ No SSR&$466$& $470$ &$466$&$480$\\ 
			&$Q_1$ No SSR&$507$& $638$ &$638$&$638$\\
			&$Q_2$ No SSR&$514$& $638$ &$638$&$638$\\
			&$Q_3$ No SSR&$638$& $638$ &$638$&$638$\\
			&$Q_4$ No SSR&$638$& $638$ &$638$&$638$\\
			&ASS No SSR&$560$($64$)& $625$($41$) &$635$($20$)&$638$($9$)\\
			\cline{2-6}
			&Power SSR&$2.44\%$& $73.21\%$ &$94.21\%$&$98.62\%$\\
			&$Q_0$ SSR&$466$&$463$  &$448$&$447$\\ 
			&$Q_1$ SSR&$507$& $503$ &$492$&$487$\\
			&$Q_2$ SSR&$514$& $817$ &$507$&$495$\\
			&$Q_3$ SSR&$1276$& $1276$ &$903$&$513$\\
			&$Q_4$ SSR&$1276$& $1276$ &$1202$&$1196$\\
			&ASS SSR&$782$($358$)& $880$($349$) &$711$($308$)&$589$($227$)\\
			&$\%$ Rejected with SSR, not without&$3.0\%$& $60.1\%$ &$80.8\%$&$89.1\%$\\
			&$\%$ Not rejected with SSR, rejected without&$48.8\%$& $7.4\%$ &$2.6\%$&$0.9\%$\\
			\hline
			\textbf{Standard CP}&Average $\#$ Days&$1032$&$1032$ &$1032$&$1032$\\
			&FS&$60.1\%$&$9.7\%$ &$2.1\%$&$0.3\%$\\
			&Power Loss&$0.09\%$& $0.93\%$&$0.67\%$&$0.18\%$\\
			&Power No SSR&$2.53\%$& $57.74\%$ &$89.85\%$&$98.55\%$\\
			&$Q_0$ No SSR&$536$& $540$ &$540$&$539$\\ 
			&$Q_1$ No SSR&$540$& $638$ &$638$&$638$\\
			&$Q_2$ No SSR&$543$& $638$ &$638$&$638$\\
			&$Q_3$ No SSR&$638$& $638$ &$638$&$638$\\
			&$Q_4$ No SSR&$638$& $638$ &$638$&$638$\\
			&ASS No SSR&$579$($48$)& $629$($29$) &$636$($14$)&$638$($5$)\\
			\cline{2-6} 
			&Power SSR&$2.52\%$& $74.50\%$ &$95.05\%$&$99.00\%$\\
			&$Q_0$ SSR&$536$&$536$  &$536$&$535$\\ 
			&$Q_1$ SSR&$540$& $540$ &$540$&$539$\\
			&$Q_2$ SSR&$543$& $827$ &$543$&$542$\\
			&$Q_3$ SSR&$1276$& $1276$ &$914$&$546$\\
			&$Q_4$ SSR&$1276$& $1276$ &$1205$&$1199$\\
			&ASS SSR&$802$($345$)& $898$($334$) &$735$($291$)&$627$($208$)\\
			&$\%$ Rejected with SSR, not without&$3.2\%$& $61.3\%$ &$80.5\%$&$82.5\%$\\
			&$\%$ Not rejected with SSR, rejected without&$46.9\%$& $5.5\%$ &$1.4\%$&$0.5\%$\\  \hline				
			\hline
			&&  &&&
		\end{tabular}
		\begin{tablenotes}
			\small
			\item Note: Power One Stage Trial, a reference power of one stage trial where no interim analyses or adaptations are conducted; Average $\#$ Days, average number of days elapsed since beginning of the study; FS, probability to stop for futility using O'Brien Fleming boundaries; Power Loss, loss of power when conducting an interim analysis compared to the design power; Power SSR/No SSR, Power when there is/is no sample size re-assessment; $Q_i$ SSR/No SSR, $i^{th}$ quantile of the sample size when there is/is no sample size re-assessment; ASS SSR/No SSR, average sample size over both stages and its standard deviation (in brackets); $\%$ Rejected with SSR, not without, proportion of trials that were not stopped and not rejected without SSR that are rejected with SSR; $\%$ Not rejected with SSR, rejected without, proportion of trials that were not stopped and rejected without SSR that are not rejected with SSR.			
		\end{tablenotes}
	\end{threeparttable}
\end{table}

\end{document}